\documentclass[traditabstract]{aa}

\usepackage[nonamebreak]{natbib}
\usepackage[stable]{footmisc}
\usepackage{amsmath}
\usepackage[varg]{txfonts}
\usepackage{placeins}
\usepackage{natbib}
\bibpunct{(}{)}{;}{a}{}{,}
\usepackage{graphicx}
\usepackage{ifthen}
\usepackage{mathtools}
\usepackage[breaklinks, colorlinks, citecolor=blue]{hyperref}
\usepackage{fixltx2e}
\usepackage{url}
\usepackage[table,usenames,dvipsnames]{xcolor}
\usepackage{comment} 
\usepackage{datetime} 
\usepackage{soul,ulem} 
\usepackage{lscape}

\def\setsymbol#1#2{\expandafter\def\csname #1\endcsname{#2}}
\def\getsymbol#1{\csname #1\endcsname}

\def\Planck{\textit{Planck}}

\newbox\tablebox    \newdimen\tablewidth
\def\leaderfil{\leaders\hbox to 5pt{\hss.\hss}\hfil}
\def\endPlancktable{\tablewidth=\columnwidth 
    $$\hss\copy\tablebox\hss$$
    \vskip-\lastskip\vskip -2pt}
\def\endPlancktablewide{\tablewidth=\textwidth 
    $$\hss\copy\tablebox\hss$$
    \vskip-\lastskip\vskip -2pt}
\def\tablenote#1 #2\par{\begingroup \parindent=0.8em
    \abovedisplayshortskip=0pt\belowdisplayshortskip=0pt
    \noindent
    $$\hss\vbox{\hsize\tablewidth \hangindent=\parindent \hangafter=1 \noindent
    \hbox to \parindent{$^#1$\hss}\strut#2\strut\par}\hss$$
    \endgroup}
\def\doubleline{\vskip 3pt\hrule \vskip 1.5pt \hrule \vskip 5pt}

\def\L2{\ifmmode L_2\else $L_2$\fi}

\def\DeltaT{\ifmmode \Delta T\else $\Delta T$\fi}
\def\deltat{\ifmmode \Delta t\else $\Delta t$\fi}
\def\fknee{\ifmmode f_{\rm knee}\else $f_{\rm knee}$\fi}
\def\Fmax{\ifmmode F_{\rm max}\else $F_{\rm max}$\fi}
\def\solar{\ifmmode{\rm M}_{\mathord\odot}\else${\rm M}_{\mathord\odot}$\fi}
\def\Msolar{\ifmmode{\rm M}_{\mathord\odot}\else${\rm M}_{\mathord\odot}$\fi}
\def\Lsolar{\ifmmode{\rm L}_{\mathord\odot}\else${\rm L}_{\mathord\odot}$\fi}
\def\inv{\ifmmode^{-1}\else$^{-1}$\fi}
\def\mo{\ifmmode^{-1}\else$^{-1}$\fi}
\def\sup#1{\ifmmode ^{\rm #1}\else $^{\rm #1}$\fi}
\def\expo#1{\ifmmode \times 10^{#1}\else $\times 10^{#1}$\fi}
\def\,{\thinspace}
\def\lsim{\mathrel{\raise .4ex\hbox{\rlap{$<$}\lower 1.2ex\hbox{$\sim$}}}}
\def\gsim{\mathrel{\raise .4ex\hbox{\rlap{$>$}\lower 1.2ex\hbox{$\sim$}}}}

\def\simprop{\mathrel{\raise .4ex\hbox{\rlap{$\propto$}\lower 1.2ex\hbox{$\sim$}}}}
\def\deg{\ifmmode^\circ\else$^\circ$\fi}
\def\pdeg{\ifmmode $\setbox0=\hbox{$^{\circ}$}\rlap{\hskip.11\wd0 .}$^{\circ}
          \else \setbox0=\hbox{$^{\circ}$}\rlap{\hskip.11\wd0 .}$^{\circ}$\fi}
\def\arcs{\ifmmode {^{\scriptstyle\prime\prime}}
          \else $^{\scriptstyle\prime\prime}$\fi}
\def\arcm{\ifmmode {^{\scriptstyle\prime}}
          \else $^{\scriptstyle\prime}$\fi}
\newdimen\sa  \newdimen\sb
\def\parcs{\sa=.07em \sb=.03em
     \ifmmode \hbox{\rlap{.}}^{\scriptstyle\prime\kern -\sb\prime}\hbox{\kern -\sa}
     \else \rlap{.}$^{\scriptstyle\prime\kern -\sb\prime}$\kern -\sa\fi}
\def\parcm{\sa=.08em \sb=.03em
     \ifmmode \hbox{\rlap{.}\kern\sa}^{\scriptstyle\prime}\hbox{\kern-\sb}
     \else \rlap{.}\kern\sa$^{\scriptstyle\prime}$\kern-\sb\fi}
\def\ra[#1 #2 #3.#4]{#1\sup{h}#2\sup{m}#3\sup{s}\llap.#4}
\def\dec[#1 #2 #3.#4]{#1\deg#2\arcm#3\arcs\llap.#4}
\def\deco[#1 #2 #3]{#1\deg#2\arcm#3\arcs}
\def\rra[#1 #2]{#1\sup{h}#2\sup{m}}

\def\dots{\relax\ifmmode \ldots\else $\ldots$\fi}
\def\WHzsr{\ifmmode $W\,Hz\mo\,sr\mo$\else W\,Hz\mo\,sr\mo\fi}
\def\mHz{\ifmmode $\,mHz$\else \,mHz\fi}
\def\GHz{\ifmmode $\,GHz$\else \,GHz\fi}
\def\mKs{\ifmmode $\,mK\,s$^{1/2}\else \,mK\,s$^{1/2}$\fi}
\def\muKs{\ifmmode \,\mu$K\,s$^{1/2}\else \,$\mu$K\,s$^{1/2}$\fi}
\def\muKRJs{\ifmmode \,\mu$K$_{\rm RJ}$\,s$^{1/2}\else \,$\mu$K$_{\rm RJ}$\,s$^{1/2}$\fi}
\def\muKHz{\ifmmode \,\mu$K\,Hz$^{-1/2}\else \,$\mu$K\,Hz$^{-1/2}$\fi}
\def\MJysr{\ifmmode \,$MJy\,sr\mo$\else \,MJy\,sr\mo\fi}
\def\MJysrmK{\ifmmode \,$MJy\,sr\mo$\,mK$_{\rm CMB}\mo\else \,MJy\,sr\mo\,mK$_{\rm CMB}\mo$\fi}
\def\microns{\ifmmode \,\mu$m$\else \,$\mu$m\fi}

\def\muK{\ifmmode \,\mu$K$\else \,$\mu$\hbox{K}\fi}
\def\microK{\ifmmode \,\mu$K$\else \,$\mu$\hbox{K}\fi}
\def\muW{\ifmmode \,\mu$W$\else \,$\mu$\hbox{W}\fi}
\def\kms{\ifmmode $\,km\,s$^{-1}\else \,km\,s$^{-1}$\fi}
\def\kmsMpc{\ifmmode $\,\kms\,Mpc\mo$\else \,\kms\,Mpc\mo\fi}
\providecommand{\sorthelp}[1]{}

\def\LCDM{$\Lambda$CDM}
\def\NHUNIT{\ifmmode {\rm \,cm^{-2}} \else $\rm \,cm^{-2}$ \fi} 

\def\wmap{\WMAP}
\def\aap{{A\&A}}

\def\apjs{{ApJ Supp.}}
\def\muKcmb{\ifmmode \,\mu$K$_{\rm CMB}$\else \,$\mu$K$_{\rm CMB}$\fi}
\newcommand{\planck}{\Planck}
\newcommand{\WMAP}{WMAP}

\newcommand{\OmegaM}{\ifmmode\Omega_{\rm M}\else $\Omega_{\rm M}$\fi}

\def\WMAP{{WMAP}}

\newcommand{\onesig}[1]{(68\%, \text{#1})}

\newcommand{\commander}{{\tt Commander}}

\providecommand{\Planck}{\textit{Planck}}
\providecommand{\planck}{\Planck}

\providecommand{\text}[1]{\rm{#1}}

\providecommand{\muK}{\mu\rm{K}}

\providecommand{\LCDM}{{$\rm{\Lambda CDM}$}}

\newcommand{\begm}{\begin{pmatrix}}
\newcommand{\enm}{\end{pmatrix}}

\def\pmb#1{\setbox0=\hbox{#1}%
    \kern-.025em\copy0\kern-\wd0
    \kern.05em\copy0\kern-\wd0
    \kern-.025em\raise.0433em\box0}

\def\p2Y{\;_2Y}
\def\m2Y{\;_{-2}Y}
\def\beglet{
  \addtocounter{equation}{1}%
  \setcounter{parentequation}{\value{equation}}%
  \setcounter{equation}{0}%
  \def\theequation{\arabic{parentequation}\alph{equation}}%
  \ignorespaces
}
\def\endlet{
  \setcounter{equation}{\value{parentequation}}%
  \def\theequation{\arabic{equation}}%
}
\providecommand{\beglet}{\begin{subequations}}
\providecommand{\endlet}{\end{subequations}}

\newcommand{\mksym}[1]{\ifmmode {\rm #1}\else #1\fi}

\newcommand{\lowTEB}{\mksym{{\rm lowTEB}}}
\newcommand{\lowE}{\mksym{{\rm lowE}}}

\newcommand{\lowEB}{\mksym{{\rm lowP}}}

\providecommand{\text}[1]{\rm{#1}}

\providecommand{\muK}{\mu\rm{K}}

\providecommand{\healpix}{\texttt{HEALPix}}

\providecommand{\LCDM}{{$\rm{\Lambda CDM}$}}

\newcommand\ba{\begin{eqnarray}}
\newcommand\ea{\end{eqnarray}}
\newcommand\bea{\begin{eqnarray}}
\newcommand\eea{\end{eqnarray}}

\newcommand\be{\begin{equation}}
\newcommand\ee{\end{equation}}

\newcommand{\mbf}[1]{\ensuremath{\mathbf{#1}}}

\newcommand{\srolltwo}{{\tt SRoll2}}

\newcommand{\clik}{{\tt clik}}

\newcommand{\camb}{{\tt camb}}

\newcommand{\combdata}{WMAP+LFI}
\newcommand{\mmbf}{\mbf{m}}
\newcommand{\nbf}{\mbf{n}}
\newcommand{\Tbf}{\mathbf T}
\newcommand{\Qbf}{\mathbf Q}
\newcommand{\Ubf}{\mathbf U}
\newcommand{\mbfP}{\mathbf m^\mathrm{P}}
\newcommand{\mbfPcs}{\mmbf^{\mathrm{P,fc}}}
\newcommand{\mtbfPcs}{\widetilde{\mmbf}^{\mathrm{P,fc}}}
\newcommand{\mbfcs}{\mmbf^\mathrm{fc}}

\newcommand{\N}{\mathrm N^\mathrm{P}_\mathrm{r}}
\newcommand{\Cm}{{\mathrm C}^{-1}}
\newcommand{\CmN}{\Cm \N}
\newcommand{\Var}[1]{\mathrm{Var}\left[#1\right]}
\newcommand{\tr}[1]{\mathrm{Tr}\left[#1\right]}

\begin{document}

\title{A novel CMB polarization likelihood package for large angular scales built from combined \wmap\ and \planck\ LFI legacy maps}

\date{\vglue -1.5mm \today \vglue -5mm}

\abstract{\vglue -3mm We present a CMB large-scale polarization dataset obtained by combining \wmap\ Ka, Q and V with \Planck\ 70 GHz maps. We employ the legacy frequency maps released by the \wmap\ and \Planck\ collaborations and perform our own Galactic foreground mitigation technique, which relies on \Planck\ 353 GHz for polarized dust and on \planck\ 30 GHz and \wmap\ K for polarized synchrotron. We derive a single, optimally-noise-weighted, low-residual-foreground map and the accompanying noise covariance matrix. These are shown, through $\chi^2$ analysis, to be robust over an ample collection of Galactic masks. We use this dataset, along with the \planck\ legacy \commander\ temperature solution, to build a pixel-based low-resolution CMB likelihood package, whose robustness we test extensively with the aid of simulations, finding excellent consistency. Using this likelihood package alone, we constrain the optical depth to reionazation $\tau=0.069^{+0.011}_{-0.012}$ at $68\%$ C.L., on 54\% of the sky. Adding the \Planck\ high-$\ell$ temperature and polarization legacy likelihood, the \planck\ lensing likelihood and BAO observations we find $\tau=0.0714_{-0.0096}^{+0.0087}$ in a full $\Lambda$CDM exploration. The latter bounds are slightly less constraining than those obtained employing \Planck\ HFI CMB data for large angle polarization, that only include EE correlations. Our bounds are based on a largely independent dataset that does include also TE correlations. They are generally well compatible with \Planck\ HFI preferring  slightly higher values of $\tau$.  We make the low-resolution \planck\ and \wmap\ joint dataset publicly available along with the accompanying likelihood code.}


\authorrunning{Natale et al.}

\titlerunning{Large scale CMB polarization likelihood from \wmap\ and \planck\ LFI legacy maps}

\author{\small
U.~Natale\inst{1,2}~\thanks{umberto.natale@unife.it}
\and
L.~Pagano\inst{1,2}~\thanks{luca.pagano@unife.it}
\and
M.~Lattanzi\inst{2}
\and
M.~Migliaccio\inst{3,4}
\and
L.~P.~Colombo\inst{5}
\and
A.~Gruppuso\inst{6,7}
\and
P.~Natoli\inst{1,2}
\and
G.~Polenta\inst{8}
}

\institute{\small
Dipartimento di Fisica e Scienze della Terra, Universit\`a degli Studi di Ferrara, via Giuseppe Saragat 1, 44122 Ferrara, Italy\goodbreak
\and
Istituto Nazionale di Fisica Nucleare (INFN), Sezione di Ferrara, Via Giuseppe Saragat 1, 44122 Ferrara, Italy\goodbreak
\and
Dipartimento di Fisica, Universit\`a di Roma Tor Vergata, Via della Ricerca Scientifica 1, 00133, Roma, Italy\goodbreak
\and
Istituto Nazionale di Fisica Nucleare (INFN), Sezione di Roma 2, Via della Ricerca Scientifica 1, 00133, Roma, Italy\goodbreak
\and
Dipartimento di Fisica, Universit\`a degli Studi di Milano, Via Celoria 16, 20133 Milano, Italy\goodbreak
\and
INAF - OAS Bologna, Istituto Nazionale di Astrofisica - Osservatorio di Astrofisica e Scienza dello Spazio di Bologna, via Gobetti 101, 40129 Bologna, Italy \goodbreak
\and
Istituto Nazionale di Fisica Nucleare (INFN), Sezione di Bologna, viale Berti Pichat 6/2, 40127, Bologna, Italy\goodbreak
\and
Space Science Data Center - Agenzia Spaziale Italiana, Via del Politecnico snc, 00133, Roma, Italy\goodbreak
}

\maketitle

\section{Introduction}\label{sec:intro}

Measuring the CMB polarization at large angular scales is crucial for constraining the reionization peak and in particular for the determination of the Thomson scattering optical depth to reionization $\tau$, currently the less constrained of the $\Lambda$CDM parameters. The optical depth $\tau$ is connected to the integrated amount of free electrons along the line of sight, and provides information on how and when the first stars and galaxies formed.

Remarkable advancements have been made in this field over the last 15 years. \wmap\ \citep{hinshaw2012} and \planck\ \citep{planck2016-l06} collaborations have continuously improved the quality of large scale polarization measurements, notoriously extremely tough to clean from foregrounds and instrumental systematic effects contaminations. The current most constraining dataset is provided by the \Planck\ collaboration \citep{planck2016-l01} which uses the high frequency instrument (hereafter HFI) measurements at 100 and 143~GHz. Such results for the Legacy \planck\ release are presented in \citet{planck2016-l03} and \citet{planck2016-l05}, while an improved post-\planck\ analysis is presented in \citet{Delouis:2019bub} and in \citet{Pagano:2019tci}.

These HFI-based measurements are all specifically designed to determine the reionization optical depth, and thus mainly dedicated to the characterization of the E-modes power spectrum. This approach, consistent with the corresponding  likelihood codes delivered, 
is mainly driven by the difficulty of building reliable noise covariance matrices and by the relatively high level of residual systematic effects related to dipole and foreground temperature-to-polarization leakage. Such likelihoods, despite being the most sensitive to date, do not include the TE spectrum \citet{planck2016-l05}. Furthermore, they cannot be straightforwardly adapted to handle non-rotationally invariant cosmologies and they might need tuned-up simulations for exotic models \citep[see][section 2.2.6]{planck2016-l05}. 

For the Legacy data release, together with the HFI-based likelihood, the \Planck\ collaboration also delivered a map-based likelihood employing observations of the low frequency instrument (hereafter LFI) 70~GHz channel. The sensitivity to the reionization optical depth of the LFI-based likelihood is more than a factor 2 worse with respect to the HFI-based likelihood.

The possibility of combining the WMAP and \Planck\ observations to build a ``joint'', more constraining, dataset was first explored by some of us in \citet{Lattanzi:2016dzq}, using the data available at the time. To date, however, a combined dataset using the \wmap\ and \Planck\ legacy observations of the large scale polarization is not publicly available. Aiming to fill this gap, we present here a combined real-space polarization dataset which considers jointly the \Planck\ 70~GHz channel and the \wmap\ Ka-, Q- and V-bands. Due to the aforementioned difficulty to deal with residual systematic effects in pixel space \citep{planck2016-l05}, we do not consider HFI CMB channels, such as 100 and 143~GHz. 

An important aspect of our work is that we perform an independent analysis pipeline starting from the raw maps, as delivered by the two collaborations. From there, we build polarization masks, perform foreground cleaning through a template fitting and, finally, combine the four maps in pixel domain through inverse noise weighting. The resulting dataset, despite still having an overall higher noise than the HFI-based one, allows for an independent estimation of the reionization optical depth. Moreover, being a real-space dataset, it is suitable for a number of studies not accessible to a spectrum-based likelihood (see e.g. \citet{planck2013-p09,planck2014-a18}), and it is able to explore non-rotationally invariant cosmologies. For an exhaustive review of likelihood methods see \citet{Gerbino:2019okg}.

The paper is organized as follows. We start by describing the input datasets in Sec.~\ref{sec:data}. In Sec.~\ref{sec:mask} we present the new set of masks produced for the \Planck\ and \wmap\ data. In Sec.~\ref{sec:methods}, we introduce the main algorithms used in the paper, including a detailed discussion on the role of the regularization noise used in the analysis. In Sec.~\ref{sec:compsep} we present the results of the component separation process. In Secs.~\ref{sec:spectra} and \ref{sec:likelihood} we show the angular power spectra and study the stability of the $\tau$ estimates in different masks, also performing a Monte Carlo validation. We present our constraints on the reionization optical depth $\tau$ in Sec.~\ref{sec:cosmology}, and draw our conclusions in Sec.~\ref{sec:concl}. 

\section{Datasets}\label{sec:data}
In this section we describe the large-scale \wmap\ and \planck\ polarization maps that are used to build the combined dataset. As already mentioned, we consider, as CMB channels, the $70$\GHz\ from \Planck\ LFI \citep{planck2016-l02} and the Ka, Q, and V bands from \wmap\ \citep{bennett2012}. In the case of LFI 70\GHz\, we use the full mission map after removing the bandpass and gain mismatch leakage correction maps. These maps, described in \citep{planck2016-l02}, are part of the \planck\ 2018 legacy data release, and are publicly available through the Planck Legacy Archive\footnote{\href{http://pla.esac.esa.int/pla/}{http://pla.esac.esa.int/pla/}}. For \wmap, we use the raw 9-year frequency maps, available on the Lambda archive \footnote{\href{https://lambda.gsfc.nasa.gov/product/map/dr5/m\_products.cfm}{https://lambda.gsfc.nasa.gov/product/map/dr5/m\_products.cfm}}. In principle, we could have also considered the 44\GHz\ channel from \Planck\ LFI and the W-band from \wmap\ as CMB channels. However, we have found that both these channels show excess power, of likely spurious origin, after implementing the foreground cleaning procedure described in Sec.~\ref{sec:methods}. For this reason, we have decided not to include the 44\GHz\ and W-band channels in our analysis. Note that the \planck\ and \wmap\ collaborations 
made the same choice on similar grounds \citep{planck2016-l05, page2007}.

We employ the K-band from \wmap, LFI 30\GHz\ and HFI 353\GHz\ maps from \Planck\ as tracers of Galactic foreground emission. These are used both to generate masks excluding regions dominated by Galactic emissions, and to mitigate the astrophysical foreground contamination in the remaining parts of the sky, as explained in detail in Secs.~\ref{sec:mask} and \ref{sec:methods}. At 30\GHz\ we use the full-mission, bandpass leakage-corrected map. For the 353\GHz\ channel we select a map built only from data provided by Polarization-Sensitive Bolometers (PSB) \citep{planck2016-l03}, as  done in the low-$\ell$ analysis presented in \citep{planck2016-l05}. \wmap\ K band and \planck\ 30 GHz are used as polarized synchrotron tracer, respectively for \wmap\ and \planck\ CMB channels. \Planck\ 353 GHz is used as polarized thermal dust tracer for both WMAP and \Planck.

Since we are mainly focused on the large angular scales, it appears convenient to work with low-resolution datasets. Thus all the maps of the Stokes parameters, $\mathbf{m}=[Q, U]$, describing the measured linear polarization, have been downgraded to a \healpix\ resolution of $N_{\rm side}=16$ \citep{gorski2005}, which corresponds to a pixel size of $\sim 3.7$ degrees. A smoothing kernel has been applied to the high-resolution maps prior to the downgrading, meant to avoid aliasing into the large angular scales of the high-frequency power present in the maps. The smoothing has been performed in harmonic space, using a cosine window function \citep{Benabed:2009af,planck2016-l05}. This guarantees that the signal is left unaltered on the scales of interest, {\it i.e}, up to multipoles of $\ell=N_{\rm side}=16$, while it is smoothly set to zero on smaller scales, $\ell>3\times N_{\rm side}=48$. 

The instrumental noise properties of each low-resolution map are described by an associated pixel-pixel noise covariance matrix (NCVM). For the LFI channels, the covariance matrices are presented in \citet{planck2016-l02}. The 70\GHz\ covariance matrix has been rescaled in harmonic space in order to match the noise level of the half-difference of half-ring maps, following the procedure described in \citet{planck2016-l05}. For the HFI 353\GHz\ NCVM, we use a downgraded version of the map-making covariance matrix, which is instead generated at the native high-resolution of $N_{\rm side}=2048$. This NCVM only accounts for Q and U correlations within the same pixel, while correlations between different pixels are ignored. Finally, for \wmap\ we build the NCVMs starting from the polarization pixel-pixel inverse covariance matrices at $N_{\rm side}=16$ (Res 4) delivered by the \wmap\ team and described in \citep{page2007, bennett2012}. The cosine window function apodization is performed in harmonic space on the eigenvectors of these low resolution matrices. It is worth commenting that although exchanging the order of the smoothing and downgrading operations is clearly not an option at the map level, due to the possible presence of sub-pixel structure, it can still be acceptable for the NCVMs. 

Since all the (Q,U) NCVMs have been convolved with a smoothing function, we add to them a white noise covariance matrix with $\sigma^2 = (20\, {\rm nK})^2$, in order to guarantee that they are numerically well conditioned, as in \citet{planck2016-l05}. For consistency, noise with the same statistical properties has to be added to the corresponding maps. However, instead than adding a single noise realization to each smoothed data map, as in \citet{planck2016-l05}, we follow a different procedure, described in Sec.~\ref{sec:methods}. This ensures that our results are not biased by a particular realization of the regularization noise.

For what concerns the temperature ($T$) map, we always employ the \Planck\ 2018 \commander\ solution \citep{planck2016-l04} outside its confidence mask, which leaves available $86\%$ of the sky. This map has been filtered with a Gaussian beam of FWHM $440$ arcmin and downgraded to $N_{\rm side}=16$. Since it is reasonable to assume that the temperature noise at large angular scales is negligible, 
we only need to include the regularization noise. Thus we model the temperature NCVM as a white noise covariance matrix with $\sigma^2=(2\, \mu{\rm K})^2$, as in \citet{planck2016-l05}. We consistently handle such regularization noise following the same procedure adopted for polarization. Finally, when building the NCVM of the full TQU maps, we neglect the correlation between temperature and polarization and set to zero the corresponding off-diagonal blocks in the covariance matrix \citep{planck2014-a13}.

\section{Polarization masks}\label{sec:mask}

In order to efficiently perform the foreground cleaning and the cosmological parameter estimation, we must remove from the analysis the pixels of the data map that are most affected by foreground contamination. In temperature, we always use the \commander\ 2018 confidence mask \citep{planck2016-l05} provided by the \Planck\ collaboration. In this section we describe how the polarization masks are produced.

\begin{figure}[h]
	\hspace{-0.3cm}\includegraphics[width=0.5\textwidth]{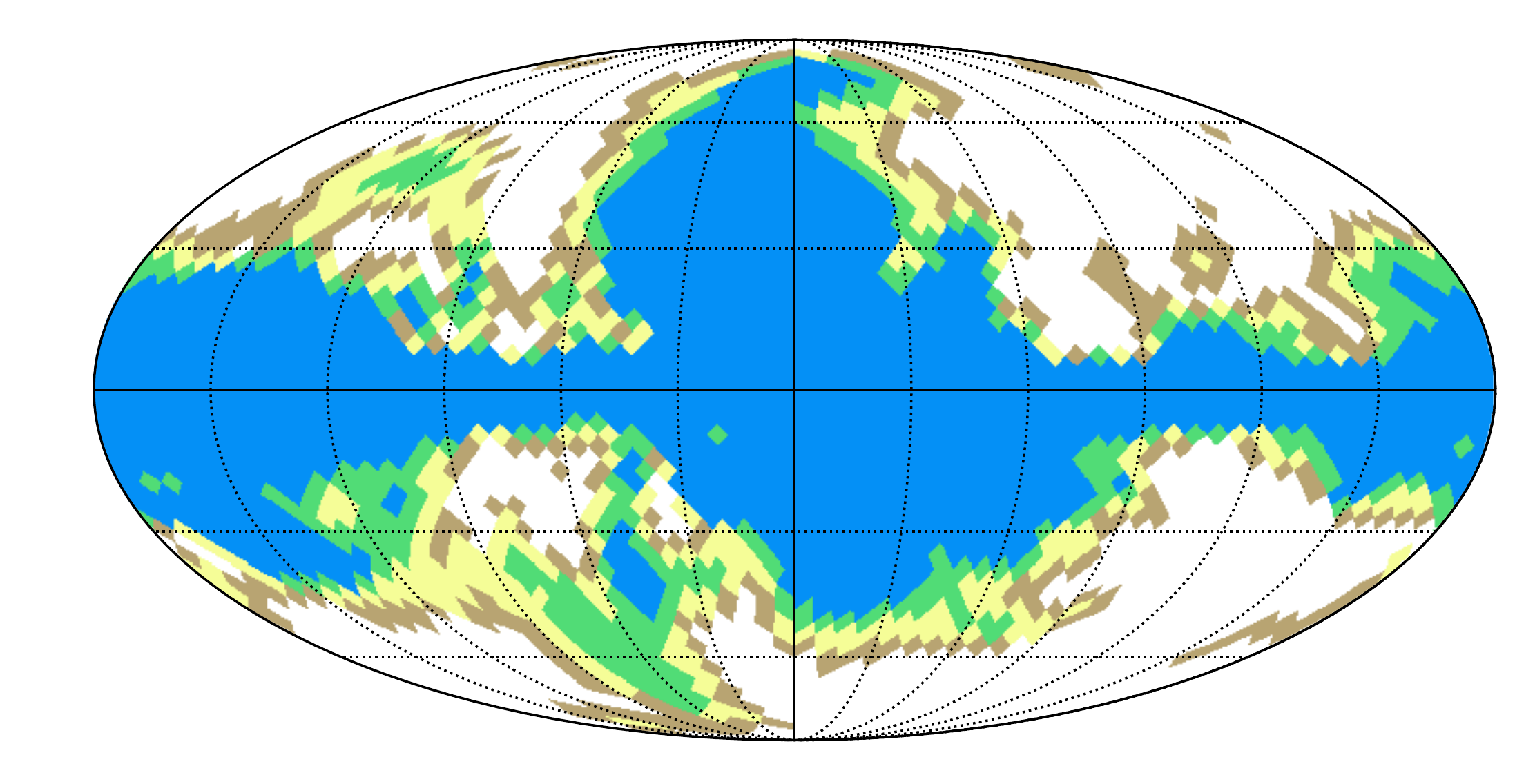}
	\caption{Subset of the masks used in the analysis of the LFI 70\GHz\ data. Values of the available sky fraction $f_\textrm{sky}$ in each mask are 30\%, 40\%, 50\% and 60\%. \label{fig:LFI_masks}}
\end{figure}

In polarization, we build two different sets of masks for \wmap\ and LFI. For LFI 70\GHz\, we use the 30 and the 353\GHz\ maps as, respectively, synchrotron (s) and dust (d) tracers, in analogy to what is done in \citet{planck2016-l05}. We first apply a Gaussian smoothing with a Full Width Half Maximum (hereafter FWHM) of $7.5^\circ$ to these input maps, taken at their native resolution of $N_{\rm side}=1024$ (30\GHz) and $N_{\rm side}=2048$ (353\GHz). Then we build maps of the polarization amplitude $P_s=\alpha\sqrt{Q_s^2+U_s^2}$ and $P_d=\beta\sqrt{Q^2_d+U^2_d}$ , where the scaling coefficients are set to $\alpha=0.063$ and $\beta=0.0077$ as estimated in \citet{planck2014-a13}. These two maps are subsequently downgraded to the \healpix\ resolution $N_{\rm side}=16$. From these maps, two separate sets of masks for synchrotron and dust emission are built as follows. We exclude pixels where the relevant polarization intensity, $P_s$ or $P_d$, is greater than a given threshold. This threshold is expressed in terms of excess intensity with respect to the corresponding mean value $\langle P_s\rangle$ or $\langle P_d\rangle$ over the whole sky. Any pair of synchrotron and dust masks can then be combined to yield a single foreground mask. Varying the threshold, we are able to build foreground masks keeping a chosen fraction of the sky. We choose to build for LFI 70\GHz\ nine different masks with equally spaced sky fractions $f_\mathrm{sky}= 30\%,\, 35\%,\,\dots,\,65\%,\, 70\%$. We do not consider larger sky fractions because, as we shall see in Sec.~\ref{sec:compsep}, we find indication of excess residual power in the LFI maps after foreground removal for masks with $f_\mathrm{sky} > 60\%$. A subset of the LFI masks is shown in Fig.~\ref{fig:LFI_masks}.
\begin{table}[h!]
	\begingroup
	\caption{Foreground scalings coefficients from \wmap\ K-band ($\alpha$) and \Planck\ 353\GHz\ ($\beta$) to the indicated \wmap\ channels.}
	\label{tab:scalings_mask_wmap}
	\nointerlineskip
	\vskip -3mm
	\footnotesize
	\setbox\tablebox=\vbox{
		\newdimen\digitwidth
		\setbox0=\hbox{\rm 0}
		\digitwidth=\wd0
		\catcode`*=\active
		\def*{\kern\digitwidth}
		\newdimen\signwidth
		\setbox0=\hbox{+}
		\signwidth=\wd0
		\catcode`!=\active
		\def!{\kern\signwidth}
		\halign{\hbox to 0.9in{#\leaderfil}\tabskip=1em&
			\hfil#\hfil\tabskip=10pt&
			\hfil#\hfil\tabskip=10pt&
			\hfil#\hfil\tabskip=0pt\cr
			\noalign{\doubleline}
			\omit\hfil Channel \hfil& $\alpha$ & $\beta$ \cr
			\noalign{\vskip 3pt\hrule\vskip 5pt}
			Ka band & $0.315$  & $0.0031$ \cr
			Q band &  $0.163$  & $0.0039$ \cr
			V band & $0.047$  & $0.0076$ \cr
			\noalign{\vskip 5pt\hrule\vskip 3pt}}}
	\endPlancktable
	\endgroup
\end{table}

\noindent

\begin{figure}[h]
	\hspace{-0.3cm}\includegraphics[width=0.5\textwidth]{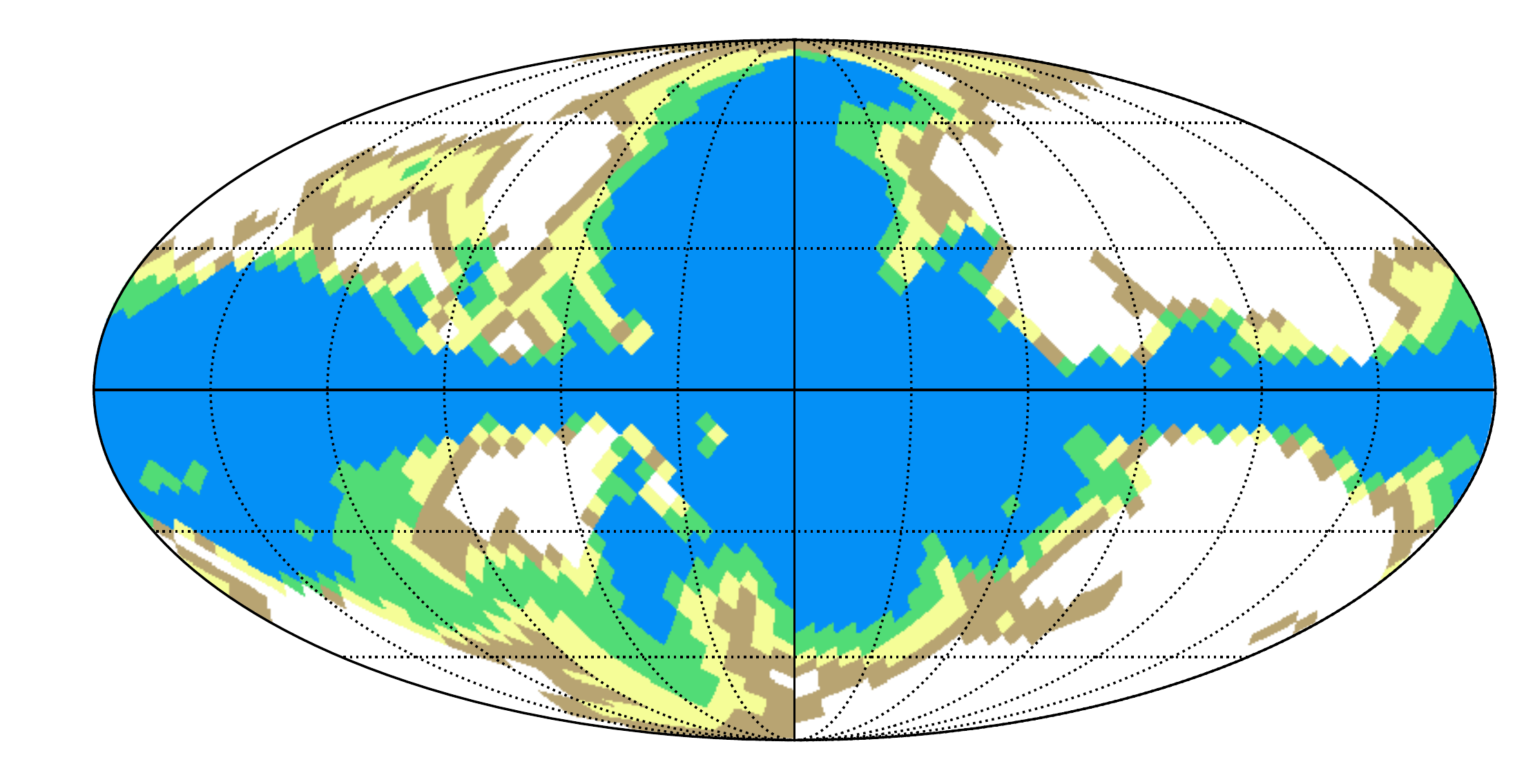}
	\caption{Subset of the masks used in the analysis of the \wmap\ data. Values of the available sky fraction $f_\textrm{sky}$ in each mask are 30\%, 40\%, 50\% and 60\%. \label{fig:WMAP_masks}}
\end{figure}

A corresponding set of masks for \wmap\ channels is built through a similar procedure. Now we use the \wmap\ K-band as a tracer for synchrotron emission and \Planck\ 353\GHz\ for dust. These are rescaled using the coefficients in \citet{Lattanzi:2016dzq}; for completeness, these values are also reported in Tab. \ref{tab:scalings_mask_wmap} here. In this case the mask structure at intermediate and high latitudes is dominated  by the synchrotron emission (i.e., by the K-band). Thus we decide to adopt the same mask for the three \wmap\ bands. This leads to a single set of ten masks, with a sky fraction ranging  from 30\% to 75\% in steps of $5\%$. A subset of the masks is shown in Fig.~\ref{fig:WMAP_masks}. 

Finally, with the aim of building a \wmap-\Planck\ LFI combined dataset, we also produce another set of masks to be used in the analysis of the joint dataset. These are built by combining pairs of \wmap\ and LFI masks, taking the pixels that are left available in at least one of the two masks. In other words, if we think of a mask as the set of all pixels that can be used in the analysis, the ``joint'' masks are the union (in the set-theory meaning of the word) of the individual \wmap\ and LFI masks. For this reason, the sky fraction of each combined mask is always equal or larger than the sky fractions of the individual masks it is built from. For example, the union of the \wmap\ and \Planck\ LFI 30\% masks has $f_\mathrm{sky} \simeq 35\%$. We then choose to produce a set of ten masks built as follows. The first seven masks are the union of each pair of \wmap\ and \Planck\ masks with the same sky fraction $f_\mathrm{sky}=30\%,\,35\%,\, \dots,\,55\%,60\%$. The remaining three masks are the union of the LFI $60\%$ mask with the \wmap\ 65\%,~70\% and 75\% masks.
The reason behind this choice is that, as mentioned above and discussed in more detail in Sec.~\ref{sec:spectra}, we do not consider the LFI masks with $f_\mathrm{sky} > 60\%$ suitable for cosmological analyses. The sky fractions for the set of union masks, together with the individual masks used to produce them, are summarized in Table~\ref{tab:union_masks}.

\begin{table}[h!]
	\begingroup
	\caption{Masks used in the analysis of the joint \wmap-\Planck\ dataset. Each mask is built as the union of the individual masks reported in the left column, and leaves the sky fraction reported on the right available for analysis.}
	\label{tab:union_masks}
	\nointerlineskip
	\vskip -3mm
	\footnotesize
	\setbox\tablebox=\vbox{
		\newdimen\digitwidth
		\setbox0=\hbox{\rm 0}
		\digitwidth=\wd0
		\catcode`*=\active
		\def*{\kern\digitwidth}
		\newdimen\signwidth
		\setbox0=\hbox{+}
		\signwidth=\wd0
		\catcode`!=\active
		\def!{\kern\signwidth}
		\halign{\hbox to 0.9in{#\leaderfil}\tabskip=1em&
			\hfil#\hfil\tabskip=10pt&
			\hfil#\hfil\tabskip=10pt&
			\hfil#\hfil\tabskip=0pt\cr
			\noalign{\doubleline}
			\omit\hfil Individual $f_\mathrm{sky}$ (\WMAP\ $\times$ \Planck\ LFI) \hfil& total $f_\mathrm{sky}$ \cr
			\noalign{\vskip 3pt\hrule\vskip 5pt}
			$30\% \times 30\%$ & $35\%$  \cr
			$35\% \times 35\%$ & $40\%$  \cr
			$40\% \times 40\%$ & $45\%$  \cr
			$45\% \times 45\%$ & $50\%$  \cr
			$50\% \times 50\%$ & $54\%$  \cr
			$55\% \times 55\%$ & $59\%$  \cr
			$60\% \times 60\%$ & $63\%$  \cr
			$65\% \times 60\%$ & $66\%$  \cr
			$70\% \times 60\%$ & $70\%$  \cr
			$75\% \times 60\%$ & $75\%$  \cr
			\noalign{\vskip 5pt\hrule\vskip 3pt}}}
	\endPlancktable
	\endgroup
\end{table}

%

\section{Methods}\label{sec:methods}

In this section, we describe the cleaning procedure and the likelihood approximation used in cosmological parameter estimation.
We give particular attention to the impact of regularization noise on both scalings and cosmological parameters estimation, and, at the end of the section, discuss how it can be mitigated.

The cleaning procedure adopted here is based on fitting foreground templates at the map level, see e.g. \citet{page2007,planck2014-a13,planck2016-l05}. Denoting the linear polarization map at a given frequency $\nu$ with $\mbfP_\nu = \left[\Qbf_\nu,\,\Ubf_\nu\right]$,
the corresponding foreground-cleaned map $\mtbfPcs_\nu$ is\footnote{Here ``fc'' stands for ``foreground-cleaned''.}:
\begin{equation}\label{eqn:map_template}
\mtbfPcs_\nu=\frac{\mbfP_\nu-\alpha_\nu\textbf{t}^\textrm{s}-\beta_\nu\textbf{t}^\textrm{d}}{1-\alpha_\nu-\beta_\nu}\,,
\end{equation}
where $\textbf{t}^\textrm{s}$ ($\textbf{t}^\textrm{d}$) and $\alpha_\nu$ ($\beta_\nu$) are respectively the tracers and the scaling coefficient for synchrotron (dust) emission described in Sec.~\ref{sec:data}. Thus, if $\mathrm{S}^\mathrm{P}$ and $\mathrm{N}^\mathrm{P}_\nu$ are, respectively, the signal and noise covariance matrices at frequency $\nu$, the fitted coefficients in Eq.~(\ref{eqn:map_template}) are estimated by minimization of the quantity: 
\begin{equation}\label{eqn:template-fitting_chi2}
\chi_\nu^2=\left(\mtbfPcs_\nu\right)^\mathrm{T} {\widetilde\mathrm{C}_\nu}^{-1}\mtbfPcs_\nu\,,
\end{equation}
where $\widetilde\mathrm{C}_\nu\equiv \bigg\langle \mtbfPcs_\nu \left(\mtbfPcs_\nu\right)^\top \bigg\rangle $ is the covariance matrix
\begin{equation}
\widetilde\mathrm{C}_\nu=\textrm{S}^\mathrm{P}(C^\mathrm{fid}_\ell)+\frac{\textrm{N}^\mathrm{P}_\nu+\alpha^2_\nu\textrm{N}^\textrm{s}+\beta^2_\nu\textrm{N}^\textrm{d}}{(1-\alpha_\nu-\beta_\nu)^2}\,.
\label{eqn:covmatcs}
\end{equation}
Note that $\chi_\nu^2$ is a $\chi^2$-distributed quantity if thought as a function of the map, but not as a function of the scalings. 

Here $\textrm{N}^\textrm{s}$ and $\textrm{N}^\textrm{d}$ are 
the polarization parts of the NCVMs for the foregrounds tracers. The signal covariance matrix is built as described in \citet{Tegmark:2001zv}, and assumes a fiducial power spectrum $C_\ell^\mathrm{fid}$, taken as the \Planck\ legacy best-fit \citep{planck2016-l06}. The inversion of $\widetilde\mathrm{C}_\nu$, necessary to compute the $\chi^2$ in Eq.~(\ref{eqn:template-fitting_chi2}), requires the addition of some regularization noise. In particular, we follow the approach used in the \Planck\ legacy analysis \citep{planck2014-a13,planck2016-l06} and consider $\sigma^{\,\mathrm{P}}_\mathrm{r}=20\,\textrm{nK}$ white noise in polarization. We thus sum a random white noise realization $\nbf^\mathrm{P}_\mathrm{r}$ with this amplitude to $\mtbfPcs_\nu$ and add a diagonal term $\mathrm{N}^\mathrm{P}_\mathrm{r} \equiv \left(\sigma^{\,\mathrm{P}}_\mathrm{r}\right)^2 \mathbf{I}$ to the covariance matrix (\ref{eqn:covmatcs}), and use these regularized objects to build the $\chi^2$ in Eq.~(\ref{eqn:template-fitting_chi2}).
In the following we will denote the cleaned map with regularization noise added as $\mbfPcs_\nu\equiv\mtbfPcs_\nu+\nbf^\mathrm{P}_\mathrm{r}$, and the associated covariance matrix as $\mathrm{C}_\nu \equiv \bigg\langle \mbfPcs_\nu \left(\mbfPcs_\nu\right)^\top \bigg\rangle = \widetilde\mathrm{C}_\nu + \mathrm{N}^\mathrm{P}_\mathrm{r}$.

Once $\alpha_\nu$ and $\beta_\nu$ have been estimated through this minimization procedure, we can define the cleaned data vector $\mbfcs_\nu\equiv\left[ \Tbf,\,\mbfPcs_\nu \right]$, with $\Tbf$ being the \commander\ map described in Sec.~\ref{sec:data}, and write down its likelihood function \citep{Gerbino:2019okg} $\mathcal{L}(C_\ell)\equiv P(\mbfcs_\nu|C_\ell)$:
\begin{multline}
-2 \log \mathcal{L}(C_\ell) =\log\big|\mathrm{S}(C_\ell) + \mathrm{N}^\mathrm{fc}_\nu\big| + \\
+ \left(\mbfcs_\nu \right)^\top (\mathrm{S}(C_\ell)+\mathrm{N}^\mathrm{fc}_\nu)^{-1} \mbfcs_\nu +\mathrm{const.}\, .\label{eq:multivariate_gaussian}
\end{multline}
The NCVM $\mathrm{N}^\mathrm{fc}_\nu$ used in the likelihood analysis is built as follows. The $TT$ block is consistent with the \commander\ map having only white regularization noise with rms $\sigma_\mathrm{r}^\mathrm{T} = 2\mu{\rm K}$, while the $TQ$ and $TU$ blocks are vanishing. The polarization part $\textrm{N}^\mathrm{P,fc}_\nu$ of the NCVM is instead given by
\begin{equation}\label{eqn:NCVM_compsep}
\textrm{N}^\mathrm{P,fc}_\nu=\frac{\textrm{N}^\mathrm{P}_\nu+\alpha^2_\nu\textrm{N}_{\textrm{s}}+\beta^2_\nu\textrm{N}_{\textrm{d}}+\sigma^2_{\alpha_\nu}\textbf{t}^\textrm{s}(\textbf{t}^\textrm{s})^\top+\sigma^2_{\beta_\nu}\textbf{t}^\textrm{d}(\textbf{t}^\textrm{d})^\top}{(1-\alpha_\nu-\beta_\nu)^2} + \mathrm{N}^\mathrm{P}_\mathrm{r} \, ,
\end{equation}
where $\sigma_{\alpha_\nu}$ and $\sigma_{\beta_\nu}$ are the uncertainties in the estimates of foreground scaling coefficients and $\textbf{t}^\textrm{s,d}(\textbf{t}^\textrm{s,d})^\top$ is the outer product of the tracer maps.

The addition of regularization noise has a small, but not completely negligible, impact on the determination of the foreground scaling coefficients, and, consequently, on cosmological parameter estimates. In fact, the extra noise added to the map increases the scatter of point estimates (e.g. the posterior mean) of parameter values around the true value. Moreover, the extra term added to the NCVM increases parameter uncertainties. In what follows, we first assess the magnitude of the former effect at the level of both scaling coefficients and cosmological parameters. We then illustrate how we manage to avoid extracting a particular noise realization which leads to nonnegligible scatter (as compared to the one caused by instrumental noise).

In order to show and quantify the extra scatter in the estimates of $\alpha$, $\beta$ and cosmological parameters induced by regolarization noise, we proceed as follows.
We draw 1000 white noise realizations $\nbf_{\mathrm{r},i}$ ($i=1,\ldots,1000$) with rms of $2\,\mu\mathrm{K}$ in temperature and $20\,\mathrm{nK}$ in polarization. We then estimate $\alpha$ and $\beta$ on the \Planck\ 70 GHz channel, following the procedure illustrated above, using each of the realization just described as the regularization noise map. For the sake of this test we adopt a mask that retains 50\% of the sky.  This procedure results in 1000 Monte Carlo estimates $\alpha_i$ and $\beta_i$. Once the scaling coefficients have been obtained, we further proceed with estimation of cosmological parameters $\Big(\log(10^{10} A_s)\Big)_i$ and $\tau_i$ from the likelihood in Eq.~(\ref{eq:multivariate_gaussian}). Note that in this last step we consistently use the same regularization noise that was used when fitting the scaling coefficients.

\begin{figure}[h]
	\includegraphics[width=0.5\textwidth]{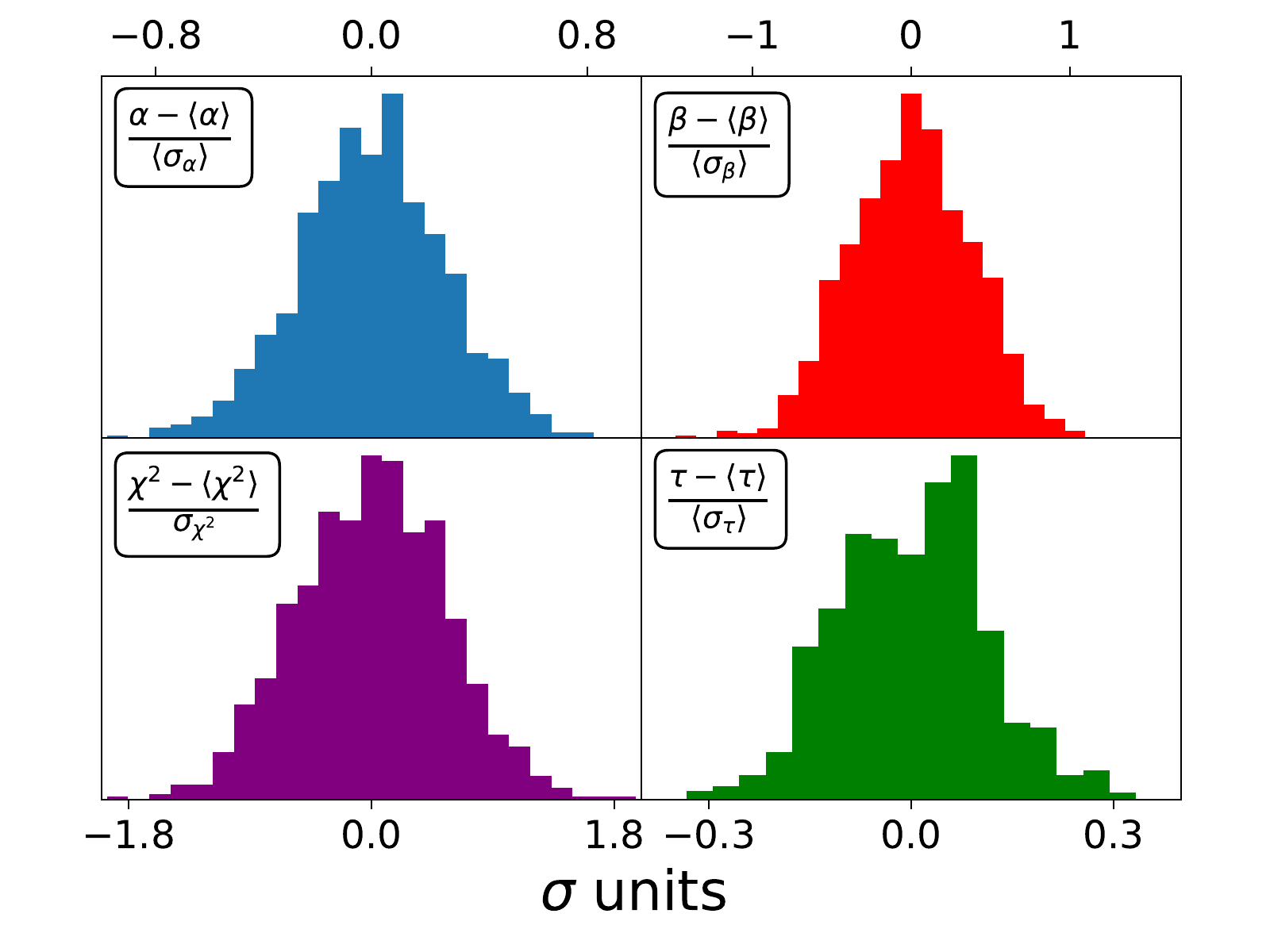}
	\caption{Histograms of the expected scatter in the recovered foreground scalings, in the $\chi^2$ of the component separation, and in the measured $\tau$ due to the regularization noise. For each quantity, we show the distance from the center of the empirical distribution in units of $\sigma$.\label{fig:realization_regularization_noise_dependence}}
\end{figure}

Since the CMB signal and the instrumental noise are the same in each map belonging to this ensemble, the scatter in the recovered values of the parameters provides an estimate of the dependence on the regularization noise realization, at the level of both scalings and cosmological parameters. The results of this procedure are shown in Fig.~\ref{fig:realization_regularization_noise_dependence}, where we show the distribution of the $\alpha_i$'s, $\beta_i$'s and $\tau_i$'s with respect to the mean value, in units of the average uncertainty. We also show the $\chi^2$ computed from Eq.~(\ref{eqn:template-fitting_chi2}), in units of $\sigma_{\chi^2} = \sqrt{2 N_\mathrm{dof}}$.

We then compute their standard deviation, i.e. $\sqrt{\langle (\theta- \langle \theta\rangle)^2\rangle}/\langle\sigma_\theta\rangle$, where $\theta=\left\{\alpha\,,\beta\,,\chi^{2}\,,\tau\right\}$. For the synchrotron scaling coefficient $\alpha$ the scatter induced by the extra noise is, on average, 0.27 times the average parameter uncertainty. In other words, 68\% of the $\alpha_i$ deviate from $\langle \alpha \rangle$ by less than $0.27\times \langle \sigma_\alpha \rangle$. The corresponding value for $\beta$ is 0.38 times the average parameter uncertainty. The $\chi^2$ of the cleaned map is the most affected quantity by the particular realization of regularization noise. In fact, the impact is, at the $1\sigma$ level, at most 0.55 times the expected width of a $\chi^2$ distribution with $N_\mathrm{dof}$ degrees of freedom. This extra scatter in the scaling estimates induces a smaller, but still non-negligible, effect on the final $\tau$ determination. The effect on $\tau$, at one standard deviation of the distribution, equals 11\% of its average uncertainty. 

Thus, when we add regularization noise, we  pay the price of an increased parameter uncertainty, and we might also be prone to unwanted parameter shifts caused by an unlucky choice of the actual noise realisation used. For example a 3-$\sigma$ noise realization can easily shift the scalings by $\sim$1 $\sigma$ and $\tau$ by $0.3$ $\sigma$. In fact roughly 1\% of the noise realizations in our Monte Carlo resulted in shifts larger than $1$ and $0.3 \sigma$'s for the scalings and $\tau$, respectively.

A possible way to avoid large parameter shifts is to somehow average over different realizations of the regularization noise. In order to do so,
we draw $N_{it}=1000$ white noise realizations $\nbf_{\mathrm{r},i}$ ($i=1,\ldots,1000$) with $20\,\mathrm{nK}$ rms. For given values of $\alpha$ and $\beta$, these are used to build as many cleaned polarization maps $\mbfPcs_i = \mtbfPcs + \nbf_{\mathrm{r},i}$ and the following quantity:
\begin{equation}
\overline{\chi^2} = \frac{1}{N_{it}}\sum_{i=1}^{N_{it}} \left(\mbfPcs_i\right)^\top  \mathrm{C}^{-1}\mbfPcs_i \, .
\label{eq:chi2bar}
\end{equation}
Note that $\overline{\chi^2}$ does not follow a chi-square distribution. It is straightforward to show that its expectation value over the regularization noise is 
\begin{equation}
\left \langle \,\overline{\chi^2} \, \right \rangle_{\textbf{n}_\textrm{r}} = \left(\mtbfPcs\right)^\top{\mathrm{C}}^{-1}\mtbfPcs+\textrm{Tr}\left(\mathrm{C}^{-1}\textrm{N}^\mathrm{P}_{\textrm{r}}\right) \, ,
\end{equation}
which is the same as the expectation value of the $\chi^2$ built from a single regularized map, $\chi^2 = \left(\mbfPcs\right)^{\top} \mathrm{C}^{-1} \mbfPcs$. Also, this expectation value is different
from the value of the $\chi^2$ on the regularization noise-free map, $\left(\mtbfPcs\right)^\mathrm{T} {\widetilde\mathrm{C}}^{-1}\mtbfPcs$.
The variance associated to $\overline{\chi^2}$ is:
\begin{align}
\Var{\, \overline{\chi^2}\,}_{\textbf{n}_\textrm{r}} = \frac{1}{N_{it}}\Bigg\{4 \left(\mtbfPcs\right)^\top \Cm \N \Cm \mtbfPcs  +2 \tr{\left( \CmN\right)^2}\Bigg\}
\end{align}
that, as should be expected, goes to $0$ as the number of noise realizations, over which the average is performed, increases.

For these reasons, we choose to minimize the quantity in Eq.~(\ref{eq:chi2bar}) to obtain estimates of the scaling coefficients 
that are less dependent on the particular realization of regularization noise. Similarly, when estimating cosmological parameters, we perform an analogous procedure, by drawing $N_{it}=1000$ noise realizations in temperature and polarization, and using the average of the quantity defined in Eq.~{\ref{eq:multivariate_gaussian}} over these realizations. The results of these procedures are presented in the next sections.

\section{Foreground cleaning}\label{sec:compsep}

In this section, we discuss the results of the estimation of the syncrotron and dust scaling coefficients for the different channels, in various masks. We also discuss how this leads to the choice of the ``confidence'' masks that are used to produce the foreground-cleaned maps for each channel, and how inverse-noise-weighted combinations of these maps are built.

We clean independently the four cosmological channels (i.e. \wmap\ Ka, Q and V bands and Planck 70\GHz), following the template-fitting procedure described in Sec.~\ref{sec:methods}. We thus minimize Eq.~(\ref{eq:chi2bar}) to estimate the synchrotron, $\alpha_\nu$, and dust, $\beta_\nu$, scaling coefficients for each map. The final polarization map, $\mtbfPcs$, and polarization noise covariance matrix, $\textrm{N}^\textrm{P,fc}$, are given by Eqs.~(\ref{eqn:map_template}) and (\ref{eqn:NCVM_compsep}).

\begin{figure}[h]
	\includegraphics[width=0.5\textwidth]{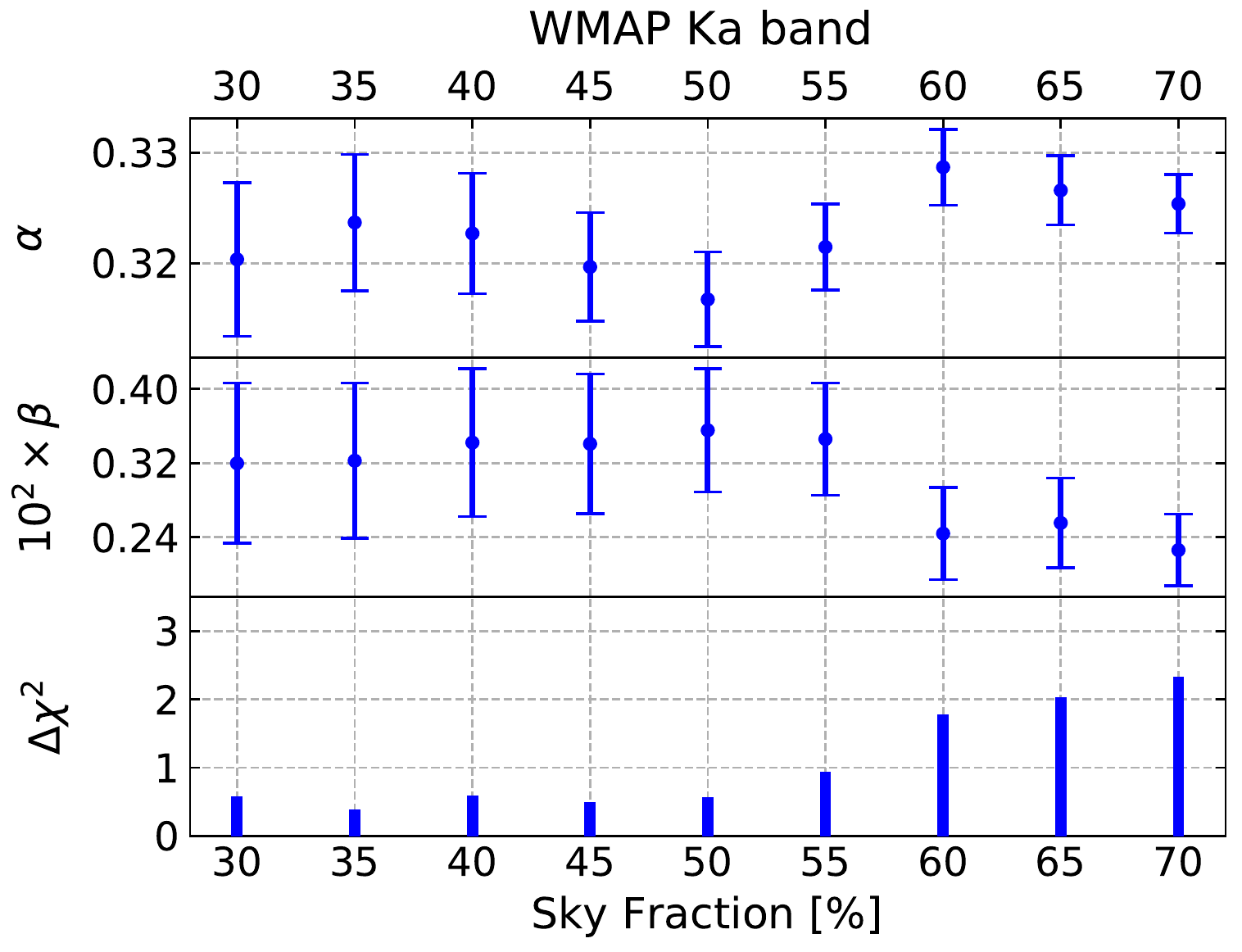}
	\caption{Scaling coefficients for synchrotron and dust (top and middle panels) estimated on different masks for \wmap\ Ka band. The bottom panel shows the excess $\chi^2$  in units of $\sqrt{2 N_{\rm dof}}$, i.e., $\Delta \chi^2 = (\chi^2 - N_\mathrm{dof})/\sqrt{2 N_{\rm dof}}$. \label{fig:scalings_Ka}}
\end{figure}

\begin{figure}[h]
	\includegraphics[width=0.5\textwidth]{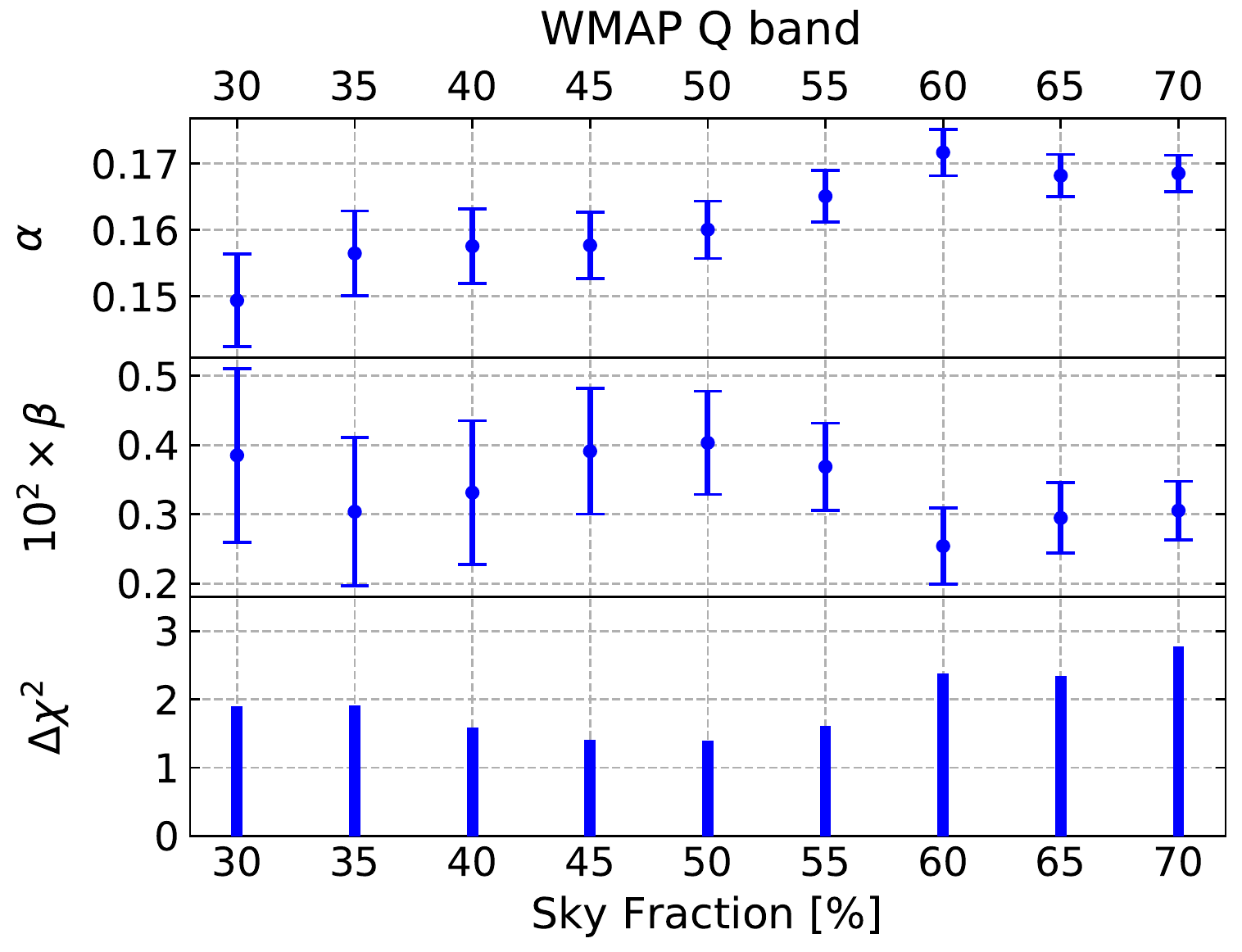}
	\caption{Same as Fig.~\ref{fig:scalings_Ka}, but for the \wmap\ Q band. \label{fig:scalings_Q}}
\end{figure}

\begin{figure}[h]
	\includegraphics[width=0.5\textwidth]{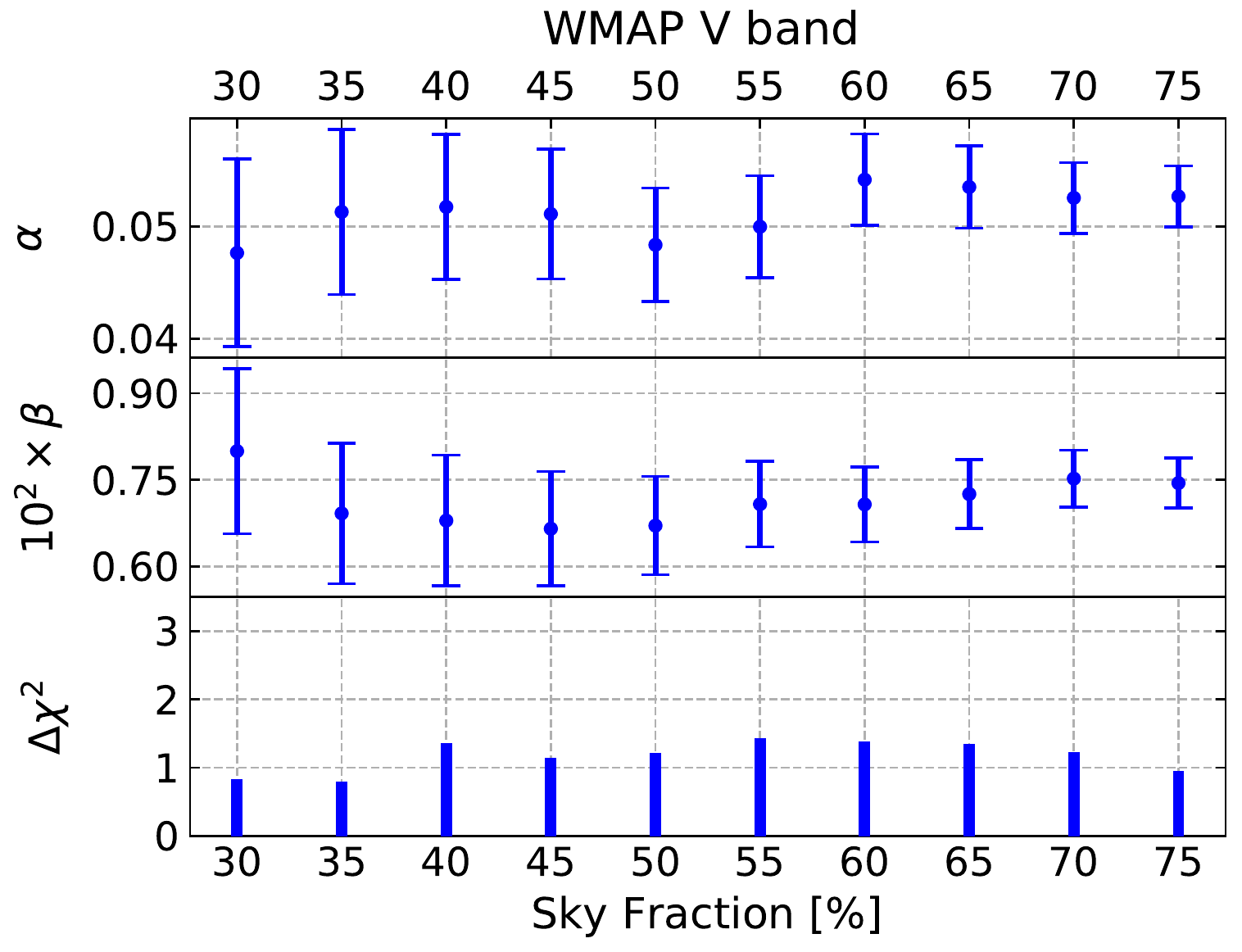}
	\caption{Same as Fig.~\ref{fig:scalings_Ka}, but for the \wmap\ V band. \label{fig:scalings_V}}
\end{figure}

\begin{figure}[h]
	\includegraphics[width=0.5\textwidth]{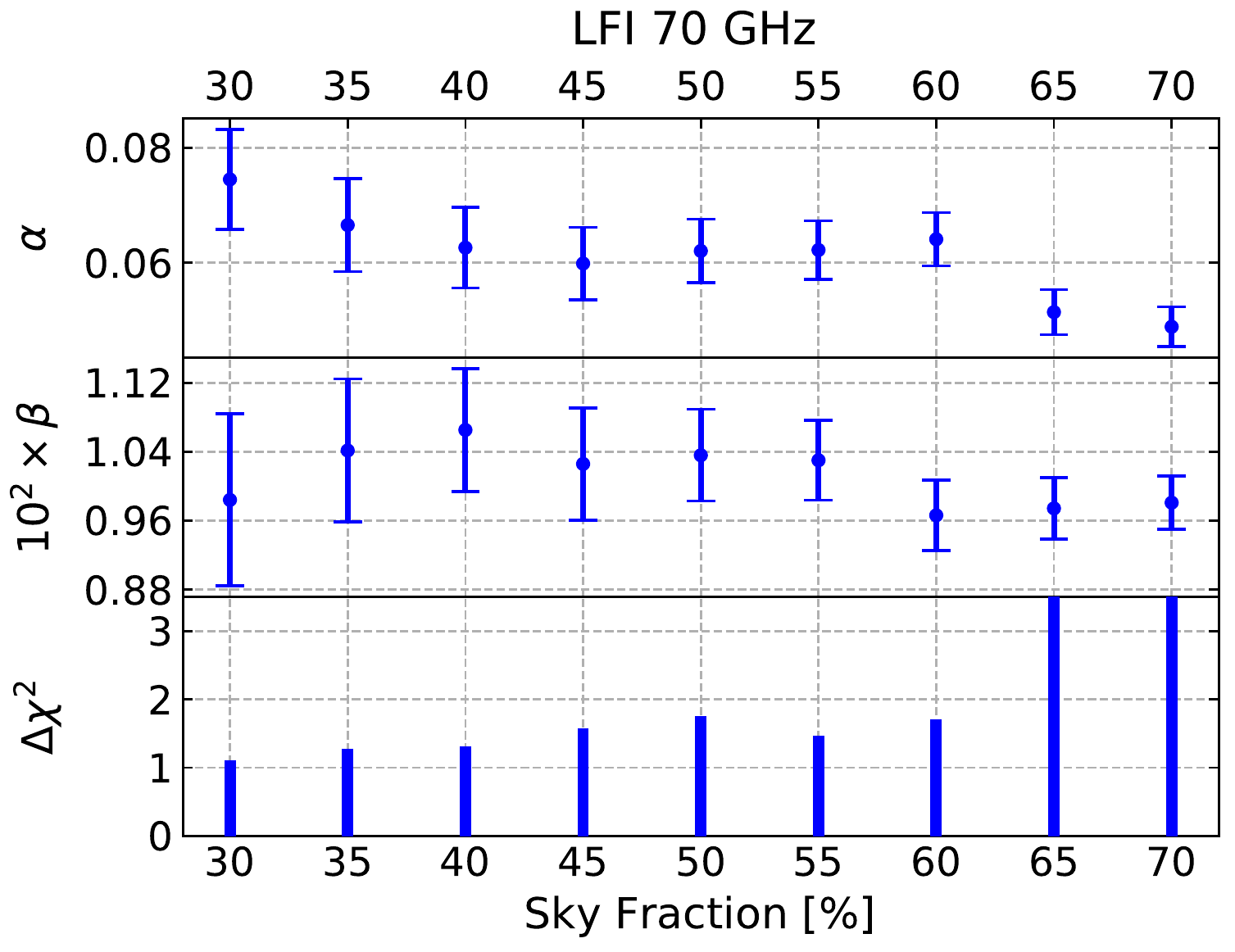}
	\caption{Same as Fig.~\ref{fig:scalings_Ka}, but for the \Planck\ LFI 70\GHz\ channel. \label{fig:scalings_70}}
\end{figure}

Figures \ref{fig:scalings_Ka}-\ref{fig:scalings_70} show the scaling coefficients computed for each cosmological channel in the masks described in Sec.~\ref{sec:mask}. In the bottom panel of each figure, we also show the excess $\chi^2$ in units of the expected dispersion, $\sqrt{2 N_{\rm dof}}$, i.e.: $\Delta \chi^2 = (\chi_\nu^2 - N_\mathrm{dof})/\sqrt{2 N_{\rm dof}}$,  where $\chi_\nu^2$ is computed from Eq.~(\ref{eqn:template-fitting_chi2}).

We use the $\Delta \chi^2$ values to select the processing mask to be used in the template fitting. The rule of thumb is to use the mask with the largest $f_\mathrm{sky}$ among those with $\Delta \chi^2\le 2$.
The only exception is represented by \wmap\ Ka band which shows a sudden change in both scalings, as well an noticeable increase in the excess $\chi^2$,  between the 55\% and 60\% masks (see Fig.~\ref{fig:scalings_Ka}). In this case we cautiously choose the to use the 55\% mask, even though the excess $\chi^2$ itself remains slightly below 2 also in the 60\% mask. Note how similar jumps between the 55\% and 60\% masks are evident also in the scalings of the \wmap\ Q band, shown in Fig.~\ref{fig:scalings_Q}. An interesting case is represented by \wmap\ V band (Fig.~\ref{fig:scalings_V}), for which the $\Delta \chi^2$ reaches a maximum in the 55\% mask before  decreasing for larger masks, without ever reaching the threshold $\Delta \chi^2 =2$. In this case we chose the 75\% mask.
The masks used in the foreground cleaning, together with the scaling coefficients obtained, are reported in Tab.~\ref{tab:scalings}. 
 
\begin{table}[h!]
\begingroup
\caption{Masks used to produce foreground-cleaned maps for each channel, and the corresponding estimates for the scaling coefficients.}
\label{tab:scalings}
\nointerlineskip
\vskip -3mm
\footnotesize
\setbox\tablebox=\vbox{
   \newdimen\digitwidth
   \setbox0=\hbox{\rm 0}
   \digitwidth=\wd0
   \catcode`*=\active
   \def*{\kern\digitwidth}
   \newdimen\signwidth
   \setbox0=\hbox{+}
   \signwidth=\wd0
   \catcode`!=\active
   \def!{\kern\signwidth}
\halign{\hbox to 0.9in{#\leaderfil}\tabskip=1em&
  \hfil#\hfil\tabskip=10pt&
  \hfil#\hfil\tabskip=10pt&
  \hfil#\hfil\tabskip=0pt\cr
\noalign{\doubleline}
\omit\hfil Channel \hfil& Mask & $\alpha$ & $\beta$ \cr
\noalign{\vskip 3pt\hrule\vskip 5pt}
Ka band & 55\% & $0.3215 \pm 0.0039$  & $0.00346 \pm 0.00061$ \cr
Q band & 55\% & $0.1651\pm 0.0039$  & $0.00369 \pm 0.00063$ \cr
V band & 75\% & $0.0527 \pm 0.0027$  & $0.00744\pm 0.00043$ \cr
70\GHz\ & 60\% & $0.0641\pm 0.0046$  & $0.00966\pm 0.00041$ \cr
\noalign{\vskip 5pt\hrule\vskip 3pt}}}
\endPlancktable
\endgroup
\end{table}

The resulting cleaned maps can then be combined together to build inverse-noise-weighted maps. In particular, we build two combinations: the first is a ``\wmap-only'' map built from the Ka, Q and V bands, while the second is a joint ``\wmap+LFI'' map, that uses the Ka, Q, V \wmap\ bands and \Planck\ LFI 70\GHz\ (hereafter \combdata). 

In more detail, if $\mtbfPcs_\nu$ is the final cleaned map of the band $\nu$ with corresponding noise-covariance matrix $\textrm{N}_{\nu}^{\textrm{P,fc}}$, the final noise weighted map $\textbf{m}^{\textrm{nw}}$ is built as:
\begin{equation}\label{eqn:inverse_noise_weighted_map}
\begin{split}
	\textbf{m}^{\textrm{nw}}&=\left[\sum_{\nu}\left(\textrm{N}_{\nu}^{\textrm{P,fc}}\right)^{-1}\right]^{-1}\sum_{\nu}\left(\textrm{N}_{\nu}^{\textrm{P,fc}}\right)^{-1}\mtbfPcs_\nu\\
&=\overline{\mathcal{N}}\sum_{\nu}\left(\textrm{N}_{\nu}^{\textrm{P,fc}}\right)^{-1}\mtbfPcs_\nu\,,
\end{split}
\end{equation}
where we define the total noise covariance matrix:
\begin{equation}
	\overline{\mathcal{N}}=\left[\sum_{\nu}\left(\textrm{N}_{\nu}^{\textrm{P,fc}}\right)^{-1}\right]^{-1}\,.
\end{equation}
Note that in Eq. (\ref{eqn:inverse_noise_weighted_map}) we are using the $\mtbfPcs_\nu$, that do not contain any regularization noise.
This is because we do not want to ``bring'' the regularization noise into the noise-weighted map; in particular we want to avoid possible biases in parameter estimates induced by particular realizations of the regularization noise, as explained in Sec.~\ref{sec:methods}.
However, we are forced to use the covariance matrices $\textrm{N}_{\nu}^{\textrm{P,fc}}$ that do include regularization noise, since otherwise we would not be able to invert them. For this reason, it is evident that $\overline{\mathcal{N}}$ would be the NCVM of a noise-weighted combination built from the (un-tilded) $\mbfPcs_\nu$, but is not the NCVM of $\textbf{m}^{\textrm{nw}}$.
The actual NCVM can be computed by rewriting Eq.~\ref{eqn:inverse_noise_weighted_map}  as
\begin{equation}
\textbf{m}^{\textrm{nw}}=\overline{\mathcal{N}}\sum_{\nu}\left(\textrm{N}_{\nu}^{\textrm{P,fc}}\right)^{-1}\left(\textbf{s}+\textbf{n}^\textrm{t}_\nu-\textbf{n}^\textrm{r}_\nu\right)\,,
\end{equation}
where $\textbf{n}^{\textrm{t}}_\nu$ denotes the total noise, i.e., the sum of instrumental and regularization noise, at each frequency. 
Taking the expectation value of $\textbf{m}^{\textrm{nw}} (\textbf{m}^{\textrm{nw}})^{\textrm{T}}$ yields
\begin{equation}
\begin{split}
&\langle \textbf{m}^{\textrm{nw}} (\textbf{m}^{\textrm{nw}})^{\textrm{T}}\rangle=\\
&=\textrm{S}+\overline{\mathcal{N}}\sum_{\nu\,\nu'}\left(\textrm{N}_{\nu}^{\textrm{P,fc}}\right)^{-1}\langle(\textbf{n}^\textrm{t}_\nu-\textbf{n}^\textrm{r}_\nu)(\textbf{n}^\textrm{t}_{\nu'}-\textbf{n}^\textrm{r}_{\nu'})^\textrm{T}\rangle\left(\textrm{N}_{\nu'}^{\textrm{P,fc}}\right)^{-1}\overline{\mathcal{N}}=\\
&=\textrm{S}+\overline{\mathcal{N}}-\left(\sigma^{\,\mathrm{P}}_{\mathrm{r}}\right)^2\overline{\mathcal{N}}\sum_{\nu}\left[\left(\textrm{N}_{\nu}^{\textrm{P,fc}}\right)^{-1}\right]^2\overline{\mathcal{N}}\,,
\end{split}
\end{equation}
where we are assuming that the regularization noise rms $\sigma^{\,\mathrm{P}}_{\mathrm{r}}$ for all the involved maps is the same. 
In the last equality, we have used the fact that $\langle \textbf{n}^\textrm{t}_\nu \textbf{n}^\textrm{r}_{\nu'}\rangle = \langle \textbf{n}^\textrm{r}_\nu \textbf{n}^\textrm{r}_{\nu'}\rangle = \left(\sigma^{\,\mathrm{P}}_{\mathrm{r}}\right)^2\mathbf{I} \delta_{\nu\nu'}$. Thus, the final noise covariance matrix of the combined instrumental noise is 
\begin{equation}\label{eq:nw_NCVM}
	\mathcal{N}\equiv\overline{\mathcal{N}}-\left(\sigma^{\,\mathrm{P}}_{\mathrm{r}}\right)^2\overline{\mathcal{N}}\sum_{\nu}\left[\left(\textrm{N}_{\nu}^{\textrm{P,fc}}\right)^{-1}\right]^2\overline{\mathcal{N}}\,.
\end{equation}

\section{Power Spectra}\label{sec:spectra}

In this section we present results for the angular power spectra of the maps described in the previous sections. In particular, we
use a QML code \citep{Tegmark:1996,Tegmark:2001zv} to extract the auto power spectra of the cleaned maps described in Sec.~\ref{sec:compsep}. In our analysis, power spectra are not directly used for the cosmological parameter extraction. We mainly use them as a probe of possible residual systematics in the maps and, consequently, for selecting the masks suitable for the likelihood analysis. The main tool for performing these consistency tests is the $\chi^2$ in harmonic space, defined as:

\begin{equation}
\chi_\textrm{h}^2  = \sum_{\ell,\ell'=2}^{\ell_\textrm{max}} (C_{\ell} - C_{\ell}^\textrm{th}) \, \tens{M}_{\ell \ell^{\prime}}^{-1} \, (C_{\ell'} - C_{\ell'}^\textrm{th})\;,\label{chi2hred}
\end{equation}

\noindent where $C_{\ell}$ is the power spectrum estimated from a given map/mask combination, $\tens{M}_{\ell \ell^{\prime}}^{-1}$ is the Fisher matrix and $C_{\ell}^\textrm{th}$ is the power spectrum of a fiducial $\Lambda$CDM model with optical depth $\tau=0.065$ and logarithmic amplitude of primordial scalar fluctuations  ${\rm{ln}}(10^{10} A_s)=3.0343$. We perform separate tests for the $TE$, $TB$, $EE$, $EB$ and $BB$ power spectra. The quantity in Eq.~(\ref{chi2hred}) can be compared to the $\chi^2$ distribution with $\ell_\textrm{max}-1$ degrees of freedom, computing the corresponding probability-to-exceed (hereafter PTE). In tables \ref{tab:PTE_LFI}, \ref{tab:PTE_WMAP} and \ref{tab:PTE_LFI_WMAP} we report the PTEs  for LFI, \wmap\ and \combdata\ for different sky fractions, corresponding to the masks presented in Sec. \ref{sec:mask}. As explained there, for the \combdata\ dataset the masks are obtained by combining the individual LFI and \wmap\ masks; we refer the reader to Tab.~\ref{tab:union_masks} for details.

 We consider here $2 \le \ell\le10$ which corresponds roughly to the multipole range affected by the reionization feature. For \wmap\ the PTEs are nicely compatible with the theoretical model for all the sky fractions considered. For the LFI dataset we see $\sim2\sigma$ deviations for BB spectrum for intermediate sky fractions ($f_\textrm{sky}=40\%$ and $f_\textrm{sky}=45\%$), fluctuations reabsorbed in larger sky fractions. In the \combdata\ dataset we do not see any failure for sky fractions below 50\%, while for larger $f_\textrm{sky}$ we see a mild failure in the TB spectrum. 

\begin{table}[h!]
	\begingroup
	\caption{Probability to exceed $\chi^2_\textrm{h}$ for LFI 70\GHz\ as a function of the sky fraction. The maximum multipole used to compute $\chi^2_\textrm{h}$ is $\ell_{\textrm{max}}=10$.}
	\label{tab:PTE_LFI}
	\nointerlineskip
	\vskip -3mm
	\footnotesize
	\setbox\tablebox=\vbox{
		\newdimen\digitwidth
		\setbox0=\hbox{\rm 0}
		\digitwidth=\wd0
		\catcode`*=\active
		\def*{\kern\digitwidth}
		\newdimen\signwidth
		\setbox0=\hbox{+}
		\signwidth=\wd0
		\catcode`!=\active
		\def!{\kern\signwidth}
		\halign{\hbox to 0.7in{#\leaderfil}\tabskip=1em&
			\hfil#\hfil\tabskip=10pt&
			\hfil#\hfil\tabskip=10pt&
			\hfil#\hfil\tabskip=10pt&
			\hfil#\hfil\tabskip=10pt&
			\hfil#\hfil\tabskip=0pt\cr
			\noalign{\doubleline}
			\omit&\multispan5\hfil PTE [\%]\hfil\cr
			\noalign{\vskip -3pt}
			\omit&\multispan5\hrulefill\cr
			\noalign{\vskip 3pt}
			\omit\hfil \hspace{4pt} Sky Fraction \hfil  & TE & EE & BB & TB & EB\cr
			\noalign{\vskip 3pt\hrule\vskip 5pt}			
			30\%  &  62.2  &  87.7  &  16.0  &  56.2  &  74.2 \cr
			35\%  &  49.9  &  75.3  &  38.5  &  32.8  &  51.6 \cr
			40\%  &  24.6  &  72.7  &  4.1  &  38.7  &  51.3 \cr
			45\%  &  15.6  &  56.2  &  5.5  &  46.7  &  60.6 \cr
			50\%  &  25.7  &  43.9  &  23.5  &  45.3  &  83.4 \cr
			55\%  &  23.1  &  34.8  &  34.8  &  23.4  &  97.4 \cr
			60\%  &  29.9  &  35.4  &  20.4  &  35.2  &  97.9 \cr			
\noalign{\vskip 5pt\hrule\vskip 3pt}}}
	\endPlancktable
	\endgroup
\end{table}

\begin{table}[h!]
	\begingroup
	\caption{Probability to exceed $\chi^2_\textrm{h}$ for \wmap\ as a function of the sky fraction. The maximum multipole used to compute $\chi^2_\textrm{h}$ is $\ell_{\textrm{max}}=10$.}
	\label{tab:PTE_WMAP}
	\nointerlineskip
	\vskip -3mm
	\footnotesize
	\setbox\tablebox=\vbox{
		\newdimen\digitwidth
		\setbox0=\hbox{\rm 0}
		\digitwidth=\wd0
		\catcode`*=\active
		\def*{\kern\digitwidth}
		\newdimen\signwidth
		\setbox0=\hbox{+}
		\signwidth=\wd0
		\catcode`!=\active
		\def!{\kern\signwidth}
		\halign{\hbox to 0.7in{#\leaderfil}\tabskip=1em&
			\hfil#\hfil\tabskip=10pt&
			\hfil#\hfil\tabskip=10pt&
			\hfil#\hfil\tabskip=10pt&
			\hfil#\hfil\tabskip=10pt&
			\hfil#\hfil\tabskip=0pt\cr
			\noalign{\doubleline}
			\omit&\multispan5\hfil PTE [\%]\hfil\cr
			\noalign{\vskip -3pt}
			\omit&\multispan5\hrulefill\cr
			\noalign{\vskip 3pt}
			\omit\hfil \hspace{4pt} Sky Fraction \hfil  & TE & EE & BB & TB & EB\cr
			\noalign{\vskip 3pt\hrule\vskip 5pt}
			30\%  &  64.3  &  71.7  &  50.3  &  59.9  &  92.6 \cr
			35\%  &  91.3  &  32.5  &  81.3  &  29.0  &  97.2 \cr
			40\%  &  84.1  &  70.8  &  79.4  &  23.2  &  90.4 \cr
			45\%  &  82.2  &  91.0  &  92.6  &  19.5  &  61.0 \cr
			50\%  &  74.8  &  73.6  &  84.5  &  22.8  &  36.2 \cr
			55\%  &  81.7  &  95.3  &  65.3  &  21.3  &  50.0 \cr
			60\%  &  70.7  &  94.2  &  58.3  &  21.1  &  83.1 \cr
			65\%  &  57.5  &  92.2  &  66.2  &  20.4  &  75.2 \cr
			70\%  &  53.2  &  85.8  &  69.7  &  34.9  &  78.3 \cr
			75\%  &  55.7  &  85.4  &  67.7  &  37.5  &  63.5 \cr
			\noalign{\vskip 5pt\hrule\vskip 3pt}}}
	\endPlancktable
	\endgroup
\end{table}

\begin{table}[h!]
	\begingroup
	\caption{Probability to exceed $\chi^2_\textrm{h}$ for the \combdata\ dataset as a function of the sky fraction. The value of maximum multipole used is fixed to $\ell_{\textrm{max}}=10$.}
	\label{tab:PTE_LFI_WMAP}
	\nointerlineskip
	\vskip -3mm
	\footnotesize
	\setbox\tablebox=\vbox{
		\newdimen\digitwidth
		\setbox0=\hbox{\rm 0}
		\digitwidth=\wd0
		\catcode`*=\active
		\def*{\kern\digitwidth}
		\newdimen\signwidth
		\setbox0=\hbox{+}
		\signwidth=\wd0
		\catcode`!=\active
		\def!{\kern\signwidth}
		\halign{\hbox to 0.7in{#\leaderfil}\tabskip=1em&
			\hfil#\hfil\tabskip=10pt&
			\hfil#\hfil\tabskip=10pt&
			\hfil#\hfil\tabskip=10pt&
			\hfil#\hfil\tabskip=10pt&
			\hfil#\hfil\tabskip=0pt\cr
			\noalign{\doubleline}
			\omit&\multispan5\hfil PTE [\%]\hfil\cr
			\noalign{\vskip -3pt}
			\omit&\multispan5\hrulefill\cr
			\noalign{\vskip 3pt}
			\omit\hfil \hspace{4pt} Sky Fraction \hfil  & TE & EE & BB & TB & EB\cr
			\noalign{\vskip 3pt\hrule\vskip 5pt}
			35\%  &  41.8  &  80.6  &  11.2  &  15.9  &  11.4 \cr
			40\%  &  55.6  &  92.5  &  25.8  &  14.7  &  35.1 \cr
			45\%  &  37.6  &  93.5  &  17.4  &  18.0  &  12.7 \cr
			50\%  &  26.7  &  86.4  &  37.9  &  11.1  &  27.8 \cr
			54\%  &  31.9  &  91.6  &  36.6  &  5.8  &  33.5 \cr
			59\%  &  40.4  &  84.3  &  25.7  &  4.9  &  22.2 \cr
			63\%  &  35.3  &  85.3  &  37.1  &  12.4  &  50.2 \cr
			66\%  &  69.0  &  98.4  &  13.2  &  38.9  &  69.1 \cr
			70\%  &  72.3  &  97.8  &  26.9  &  50.3  &  74.1 \cr
			75\%  &  77.1  &  96.4  &  30.2  &  50.3  &  80.5 \cr
			\noalign{\vskip 5pt\hrule\vskip 3pt}}}
	\endPlancktable
	\endgroup
\end{table}

Based on the PTEs results a 54\% sky fraction mask represents a robust choice for the likelihood analysis of \combdata\ dataset, see also Sec.~\ref{sec:likelihood}.

We further perform additional consistency tests for the combined dataset in the chosen mask. We compute the PTEs different choices of $\ell_{\textrm{max}}$, exploring the $\chi^2$ consistency up to $\ell=15$ and $\ell=29$. The results are reported in Tab.~\ref{tab:PTE_LFI+WMAP_lmax} and show no anomaly. Finally, we also show in Tab.~\ref{tab:PTE_LFI_WMAP_ellbyell} the $\ell$-by-$\ell$ PTEs for all the polarization power spectra computed. Also in this case we do not find any particular anomaly with only three outliers above $2.5$ $\sigma$ out of a total 140 analysed multipoles. 

\begin{table}[h!]
	\begingroup
	\caption{Probability to exceed $\chi^2_\textrm{h}$ for the combined dataset \combdata\ for different choices of $\ell_{\textrm{max}}$. Here the mask used to extract the power spectra is the combined mask with $f_{\textrm{sky}}=54\%$. }
	\label{tab:PTE_LFI+WMAP_lmax}
	\nointerlineskip
	\vskip -3mm
	\footnotesize
	\setbox\tablebox=\vbox{
		\newdimen\digitwidth
		\setbox0=\hbox{\rm 0}
		\digitwidth=\wd0
		\catcode`*=\active
		\def*{\kern\digitwidth}
		\newdimen\signwidth
		\setbox0=\hbox{+}
		\signwidth=\wd0
		\catcode`!=\active
		\def!{\kern\signwidth}
		\halign{\hbox to 0.9in{#\leaderfil}\tabskip=1em&
			\hfil#\hfil\tabskip=10pt&
			\hfil#\hfil\tabskip=10pt&
			\hfil#\hfil\tabskip=0pt\cr
			\noalign{\doubleline}
			\omit&\multispan3\hfil PTE [\%]\hfil\cr
			\noalign{\vskip -3pt}
			\omit&\multispan3\hrulefill\cr
			\noalign{\vskip 3pt}
			\omit\hfil Spectrum \hfil  &$\ell_{\textrm{max}}=10$ &$\ell_{\textrm{max}}=15$ &$\ell_{\textrm{max}}=29$\cr
			\noalign{\vskip 3pt\hrule\vskip 5pt}
			TE &  31.9 & 54.2 & 66.5\cr
			EE & 91.6& 98.4 & 98.5\cr
			BB & 36.6 & 32.8 & 14.8 \cr
			TB & 5.8 & 12.9& 32.7\cr
			EB & 33.5& 58.7& 56.2\cr
			\noalign{\vskip 5pt\hrule\vskip 3pt}}}
	\endPlancktable
	\endgroup
\end{table}

\begin{table}[h!]
	\begingroup
	\caption{Probability to exceed $\chi^2_\textrm{h}$ for the \combdata\ dataset $\ell$-by-$\ell$. Here the mask used to extract the power spectra is the combined mask with $f_{\textrm{sky}}=54\%$. }
	\label{tab:PTE_LFI_WMAP_ellbyell}
	\nointerlineskip
	\vskip -3mm
	\footnotesize
	\setbox\tablebox=\vbox{
		\newdimen\digitwidth
		\setbox0=\hbox{\rm 0}
		\digitwidth=\wd0
		\catcode`*=\active
		\def*{\kern\digitwidth}
		\newdimen\signwidth
		\setbox0=\hbox{+}
		\signwidth=\wd0
		\catcode`!=\active
		\def!{\kern\signwidth}
		\halign{\hbox to 0.8in{#\leaderfil}\tabskip=1em&
			\hfil#\hfil\tabskip=10pt&
			\hfil#\hfil\tabskip=10pt&
			\hfil#\hfil\tabskip=10pt&
			\hfil#\hfil\tabskip=10pt&
			\hfil#\hfil\tabskip=10pt&
			\hfil#\hfil\tabskip=0pt\cr
			\noalign{\doubleline}
	 		\omit&\multispan5\hfil PTE [\%]\hfil\cr
			\noalign{\vskip -3pt}
			\omit&\multispan5\hrulefill\cr
			\noalign{\vskip 3pt}
			\omit\hfil Mulitpole\hfil  & TE & EE & BB & TB & EB\cr
			\noalign{\vskip 3pt\hrule\vskip 5pt}
2  &  21.1  &  71.9  &  98.9  &  51.9  &  34.4 \cr
3  &  20.6  &  26.8  &  11.8  &  5.7  &  7.2 \cr
4  &  87.0  &  77.8  &  65.9  &  33.7  &  68.4 \cr
5  &  6.0  &  56.5  &  27.2  &  17.3  &  48.3 \cr
6  &  44.9  &  44.9  &  60.9  &  49.0  &  4.5 \cr
7  &  99.3  &  36.4  &  94.4  &  1.7  &  90.6 \cr
8  &  33.3  &  73.1  &  6.5  &  11.0  &  20.0 \cr
9  &  16.4  &  36.6  &  18.9  &  40.4  &  55.0 \cr
10  &  63.8  &  79.0  &  30.7  &  66.6  &  97.1 \cr
11  &  20.8  &  95.4  &  91.1  &  24.7  &  98.3 \cr
12  &  47.1  &  53.1  &  33.0  &  27.9  &  67.3 \cr
13  &  85.4  &  53.7  &  13.8  &  38.2  &  24.7 \cr
14  &  70.3  &  48.5  &  61.7  &  58.9  &  50.6 \cr
15  &  73.1  &  72.9  &  19.1  &  94.5  &  94.6 \cr
16  &  94.3  &  71.1  &  84.4  &  42.2  &  43.1 \cr
17  &  8.0  &  21.4  &  50.1  &  28.8  &  64.0 \cr
18  &  93.7  &  73.1  &  0.9  &  59.0  &  71.5 \cr
19  &  60.2  &  60.5  &  55.5  &  38.4  &  26.1 \cr
20  &  80.6  &  48.1  &  96.1  &  14.5  &  18.5 \cr
21  &  24.0  &  38.6  &  84.3  &  43.1  &  49.1 \cr
22  &  81.0  &  72.0  &  70.6  &  81.3  &  54.6 \cr
23  &  19.6  &  62.0  &  0.7  &  60.7  &  1.0 \cr
24  &  41.0  &  13.3  &  74.1  &  85.1  &  67.9 \cr
25  &  84.0  &  33.1  &  60.8  &  71.8  &  38.4 \cr
26  &  9.9  &  27.1  &  13.2  &  36.0  &  31.4 \cr
27  &  66.5  &  30.1  &  93.9  &  68.8  &  47.7 \cr
28  &  29.8  &  86.1  &  94.4  &  26.6  &  33.4 \cr
29  &  60.3  &  38.8  &  22.5  &  12.3  &  95.1 \cr
	\noalign{\vskip 5pt\hrule\vskip 3pt}}}
	\endPlancktable
	\endgroup
\end{table}

\begin{figure}[h]
 \includegraphics[width=0.48\textwidth]{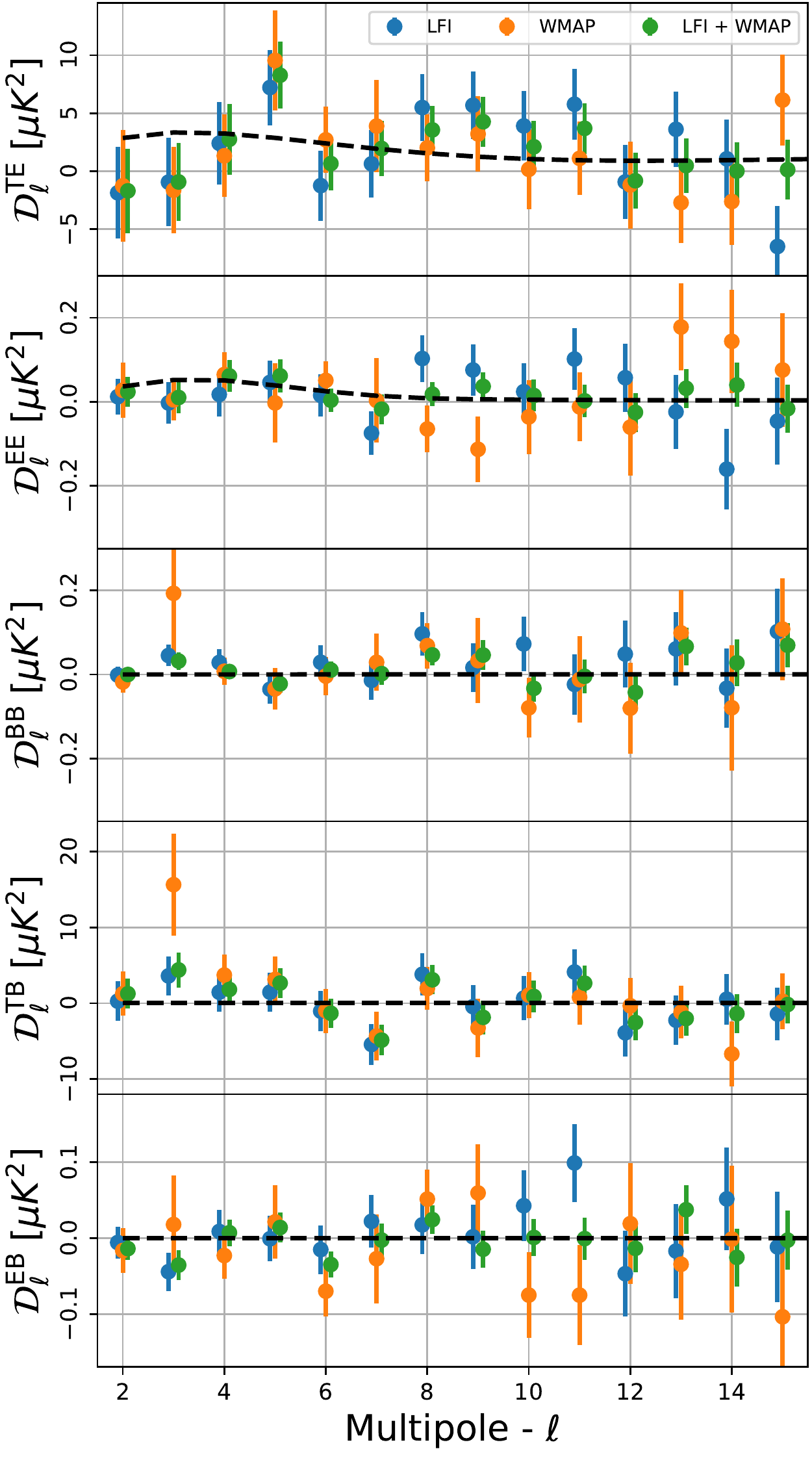}
 \caption{Polarization power spectra of the LFI 70 GHz, \wmap\ bands and \combdata. The sky fractions used are respectively 50\%, 50\% and 54\%. The dashed lines represent a $\Lambda$CDM power spectra corresponding to an optical depth value of $\tau=0.065$.\label{fig:plot_spectra_selected}}
 \end{figure}

The spectra for \wmap, LFI and \combdata\ are shown in Fig.~\ref{fig:plot_spectra_selected}, in their own 50\%, 50\% and 54\% masks, respectively. 

\section{Likelihood and validation}\label{sec:likelihood}

In this section we show the results of additional consistency tests performed at the level of parameter estimation. This allows to test and validate both the datasets produced and the likelihood algorithm.

Parameter estimates are obtained from the likelihood function in Eq.~(\ref{eq:multivariate_gaussian}). Since we are using low-resolution maps with $N_{\textrm{side}}=16$, only the $C_\ell$'s from $\ell=2$ to $\ell_{\textrm{cut}}=29$ are varied in accordance to the theoretical model that is being tested, when computing the signal covariance matrix; multipoles between $\ell_{\textrm{cut}}+1=30$ and $\ell_{\textrm{max}}=64$ are instead fixed to a fiducial $\Lambda$CDM spectrum \citep{page2007, planck2014-a13, planck2016-l05}. We follow the procedure described in Sec.~\ref{sec:methods} in order to marginalize over the regularization noise.


\begin{figure}[h]
 \includegraphics[width=0.48\textwidth]{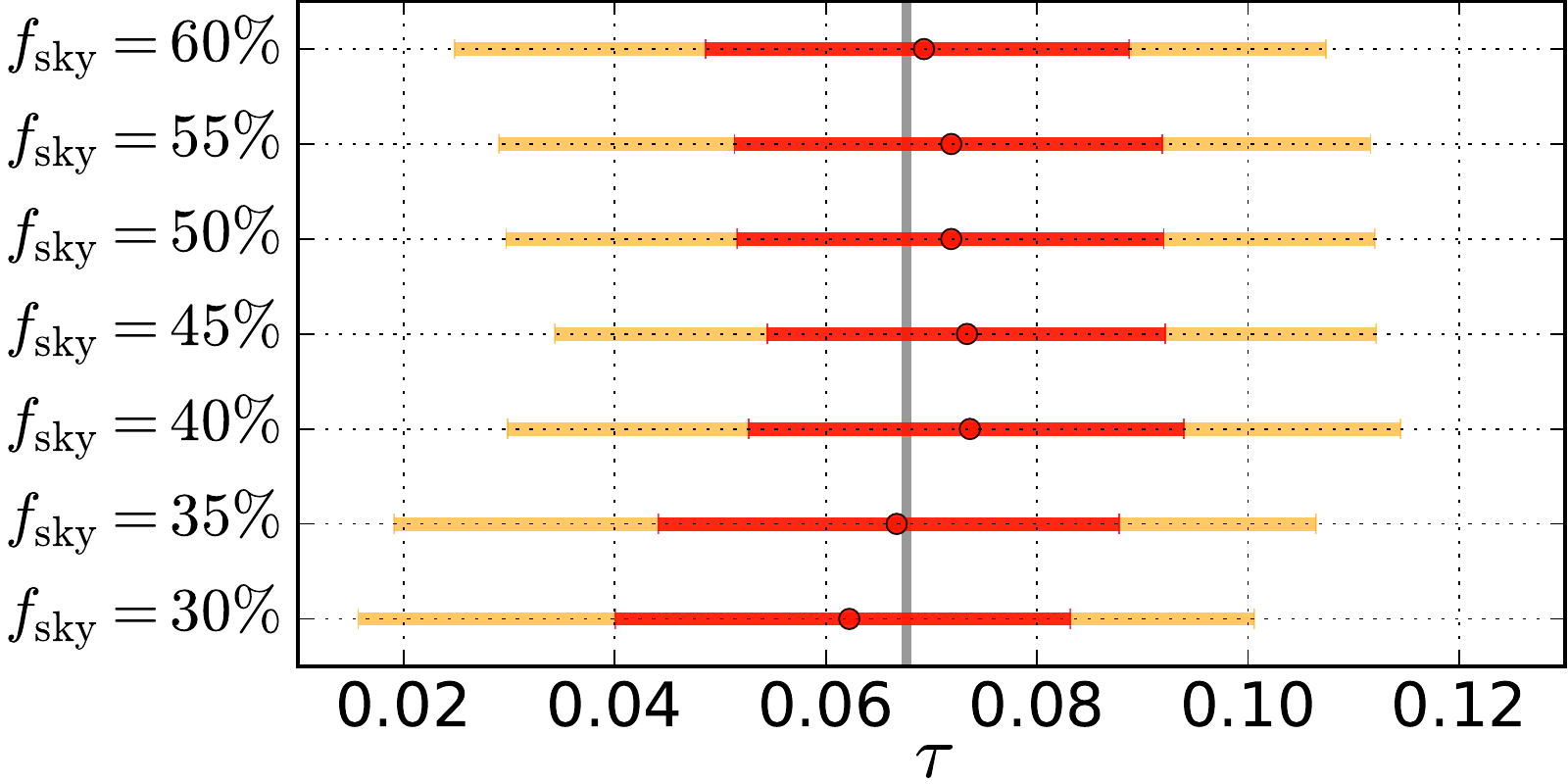}
 \caption{Estimates of $\tau$ from the LFI 70 GHz map analysed in different masks. In this and in the following plots,
 points represent best-fit values and red and yellow bars 68\% and 95\% C.L. respectively. The vertical grey line represents the best-fit value for the \combdata\ baseline dataset which uses 54\% of the sky, see text for details.\label{fig:plot_tau_fsky_Planck}}
 \end{figure}

\begin{figure}[h]
 \includegraphics[width=0.48\textwidth]{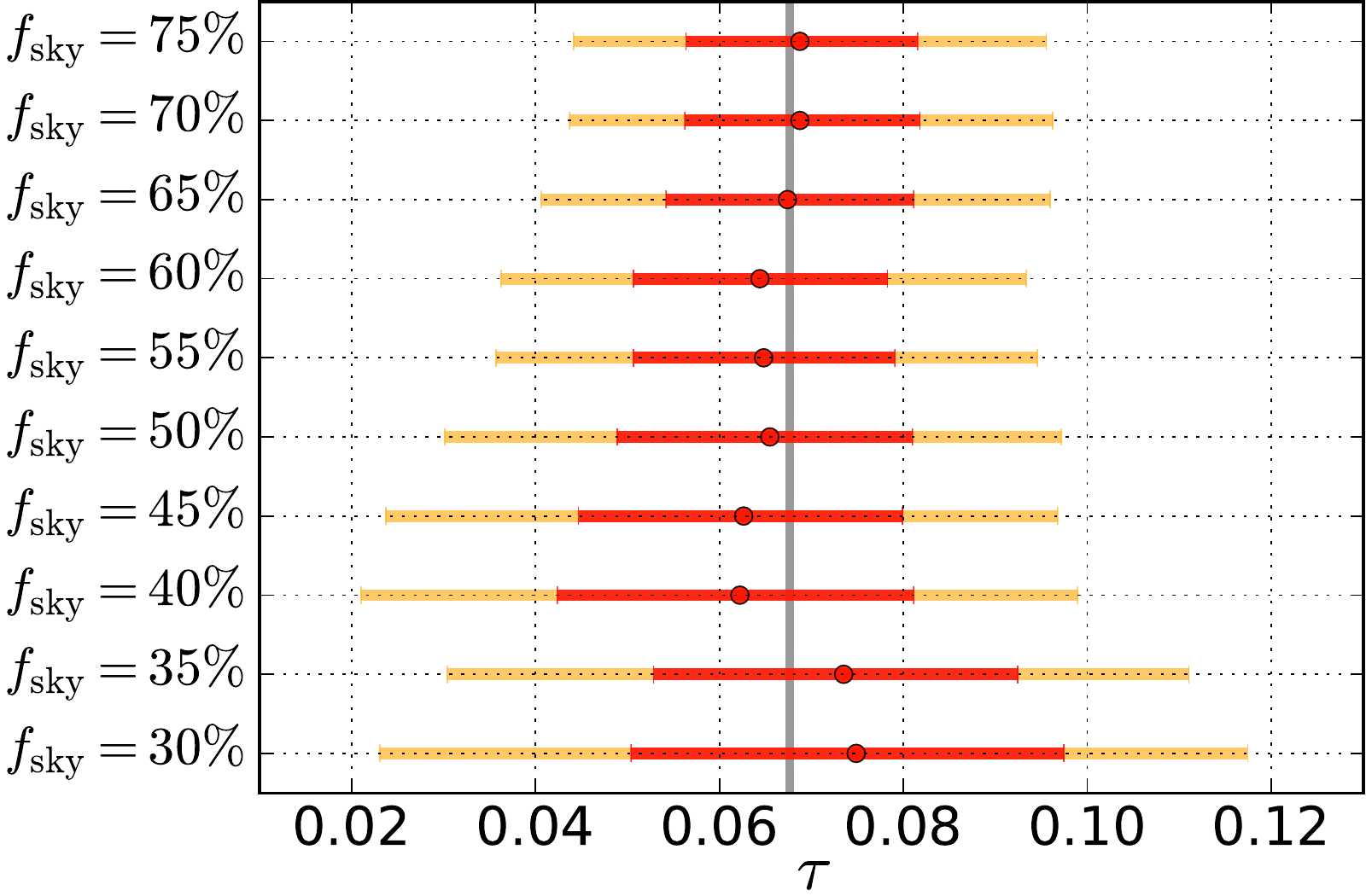}
 \caption{Same as Fig~\ref{fig:plot_tau_fsky_Planck} but for the combination of \wmap\ Ka, Q and V bands.
 \label{fig:plot_tau_fsky_WMAP}}
 \end{figure}

\begin{figure}[h!]
 \includegraphics[width=0.48\textwidth]{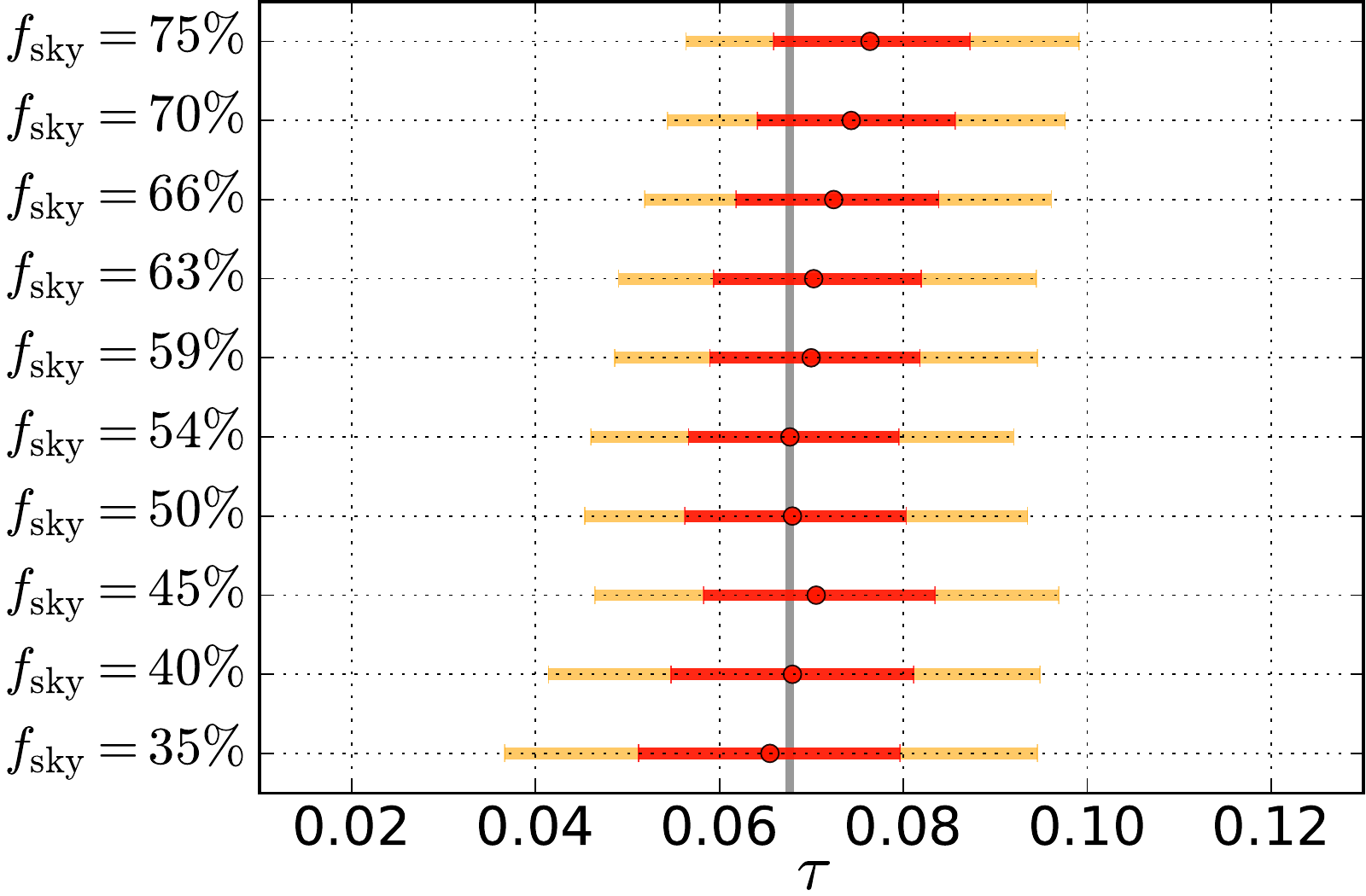}
 \caption{Same as Fig~\ref{fig:plot_tau_fsky_Planck} but for the \combdata\ dataset.
  \label{fig:plot_tau_fsky_combined}}
 \end{figure}

As consistency test for the likelihood, we explore the stability of the reionization optical depth $\tau$ constraints with respect the mask used for cosmological parameter estimation. Thus, keeping fixed the underlying datasets (i.e. map and associated covariance matrix) we only change the cosmological parameter mask used in Eq.~\ref{eq:multivariate_gaussian}. The results of these test are reported in Figures \ref{fig:plot_tau_fsky_Planck}, \ref{fig:plot_tau_fsky_WMAP} and \ref{fig:plot_tau_fsky_combined} respectively for LFI, \wmap\ and \combdata. Visually all the $\tau$ constraints are nicely compatible with each other for LFI and \wmap. For the \combdata\ the $\tau$ posteriors are still visually compatible to each other, but we observe a clear trend towards high values of $\tau$ for large sky fractions. 
It is worth to mention that all the masks we are using, for a given dataset, are nested, and largely overlapped, so relying on a simple visual comparison can be misleading and we need a more accurate statistical test to assess consistency. Thus for each dataset we generate a Monte Carlo of 1000 CMB maps with $\tau=0.065$, and a set of realistic noise simulations extracted from the noise covariance matrix of the cleaned datasets (Eq.~\ref{eqn:NCVM_compsep} and Eq~\ref{eq:nw_NCVM}) through Cholesky decomposition. 
For every mask, we process all those maps through a pixel-based likelihood algorithm implementing the function in Eq.~(\ref{eq:multivariate_gaussian}), fitting $\tau$ and $\ln(10^{10} A_{\rm s})$ from a grid of models. All the other $\Lambda$CDM parameters are kept fixed to the best-fit of \citet{Pagano:2019tci}. Then we can build the statistics
of the difference $\Delta\tau_{ij}\equiv\left|\tau_i-\tau_j\right|$ between the $\tau$ estimates for each pair of masks $\{i,\,j\}$, and finally compare this with the values of $\Delta\tau_{ij}$ obtained from the real data

In Tab. \ref{tab:pte_maskvalidation} we report the PTEs, for the three datasets analyzed, defined as the percentage of simulations that have an absolute parameter shift larger than the same quantity measured on the data. As explained above, the simulations used for this test contain only CMB signal and noise drawn by cleaned map covariance matrix. Note that  foreground residuals, and thus chance correlations between such residuals and noise realizations are not included in the simulations; this makes the test conservative since it is more difficult to pass. The scatter we see on the data is perfectly compatible with the signal plus noise simulations, independently for LFI and \wmap, and for all the sky fractions considered. The \combdata\ dataset, instead, shows mild failures for sky fractions larger than $60\%$. This comes with some surprise since the corresponding  results based on \wmap\ show excellent PTEs, see e.g. column 2 and 3 of Tab. \ref{tab:pte_maskvalidation}. Again, we have verified that the shift between the 
$\Delta\tau$ for \wmap\ and \combdata\ are compatible with what is seen in our signal plus noise simulations. We find that all the shifts are within 2-$\sigma$. Despite that, as baseline of our cosmological analysis, we opt for a conservative choice selecting the 54\% mask for \combdata. This dataset provides an error on $\tau$ 12\% smaller than \wmap\ on the 75\% mask.

\begin{table*}[htbp!]
\begingroup
\newdimen\tblskip \tblskip=5pt
\caption{Consistency of the $\tau$ parameter values estimated on different masks. For each pair of masks defined in Sect.~\ref{sec:mask}, we report the percentage of simulations with absolute parameter shift larger than the same quantity measured on the data. Each column corresponds to a different dataset, LFI (left), \wmap\ (center) and \combdata\ (right).}
\label{tab:pte_maskvalidation}
\nointerlineskip
\vskip -3mm
\setbox\tablebox=\vbox{
   \newdimen\digitwidth 
   \setbox0=\hbox{\rm 0} 
   \digitwidth=\wd0 
   \catcode`*=\active 
   \def*{\kern\digitwidth}
   \newdimen\signwidth 
   \setbox0=\hbox{+} 
   \signwidth=\wd0 
   \catcode`!=\active 
   \def!{\kern\signwidth}
\halign{\hbox to 1.0in{#\leaderfil}\tabskip=0.5em& 
  \hfil#\hfil\tabskip=1em&
    \hbox to 1.0in{#\leaderfil}\tabskip=0.5em& 
  \hfil#\hfil\tabskip=1em&
  \hbox to 1.0in{#\leaderfil}\tabskip=0.5em& 
  \hfil#\hfil\tabskip=1em\cr
\noalign{\doubleline}
\omit&\hfil LFI PTE [\%]\hfil&\omit&\hfil \wmap\ PTE [\%]\hfil&\omit&\hfil  \combdata\ PTE [\%]\hfil\cr
\noalign{\vskip -4pt}
\omit&\hrulefill&\omit&\hrulefill&\omit&\hrulefill\cr
\omit\hfil Masks\hfil & $\lvert\Delta \tau\rvert$ &\omit\hfil Masks\hfil & $\lvert\Delta \tau\rvert$ &\omit \hfil Masks\hfil & $\lvert\Delta \tau\rvert$\cr
\noalign{\vskip 3pt\hrule\vskip 5pt}
30\%$-$35\% & 49.0 & 30\%$-$35\% &  84.3 & 35\%$-$40\% & 56.9 \cr
30\%$-$40\% & 21.3 & 30\%$-$40\% &  21.1 & 35\%$-$45\% &  38.0 \cr
30\%$-$45\% & 27.2 & 30\%$-$45\% &  33.4 & 35\%$-$50\% &  72.3 \cr
30\%$-$50\% &  36.4 & 30\%$-$50\% &  49.0 & 35\%$-$54\% &  76.0 \cr
30\%$-$55\% &  40.3 & 30\%$-$55\% &  49.7 & 35\%$-$59\% &  55.9 \cr
30\%$-$60\% &  54.0 & 30\%$-$60\% &  49.6 & 35\%$-$63\% &  54.1 \cr
35\%$-$40\% &  21.2 & 30\%$-$65\% &  65.6 & 35\%$-$66\% &  37.6 \cr
35\%$-$45\% &  33.4 & 30\%$-$70\% &  71.6 & 35\%$-$70\% &  26.5 \cr
35\%$-$50\% &  52.8 & 30\%$-$75\% &  71.4 & 35\%$-$75\% &  17.6 \cr
35\%$-$55\% &  56.6 & 35\%$-$40\% &  15.9 & 40\%$-$45\% &  46.0 \cr
35\%$-$60\% &  79.6 & 35\%$-$45\% &  22.8 & 40\%$-$50\% &  97.5 \cr
40\%$-$45\% &  95.8 & 35\%$-$50\% &  43.1 & 40\%$-$54\% &  96.2 \cr
40\%$-$50\% &  75.5 & 35\%$-$55\% &  48.8 & 40\%$-$59\% &  74.9 \cr
40\%$-$55\% &  80.2 & 35\%$-$60\% &  47.3 & 40\%$-$63\% &  71.7 \cr
40\%$-$60\% &  55.1 & 35\%$-$65\% &  63.4 & 40\%$-$66\% &  48.5 \cr
45\%$-$50\% &  69.7 & 35\%$-$70\% &  72.7 & 40\%$-$70\% &  32.9 \cr
45\%$-$55\% &  76.3 & 35\%$-$75\% &  73.7 & 40\%$-$75\% &  18.6 \cr
45\%$-$60\% &  50.6 & 40\%$-$45\% &  91.5 & 45\%$-$50\% &  34.6 \cr
50\%$-$55\% &  95.6 & 40\%$-$50\% &  67.0 & 45\%$-$54\% &  49.6 \cr
50\%$-$60\% &  58.3 & 40\%$-$55\% &  79.1 & 45\%$-$59\% &  88.7 \cr
55\%$-$60\% &  40.8 & 40\%$-$60\% &  83.5 & 45\%$-$63\% &  96.8 \cr
	 &	$-$	     & 40\%$-$65\% &  64.5 & 45\%$-$66\% &  68.7 \cr
	 &	$-$	     & 40\%$-$70\% &  56.2 & 45\%$-$70\% &  46.9 \cr
	 &	$-$	     & 40\%$-$75\% &  58.1 & 45\%$-$75\% &  29.7 \cr
	 &	$-$	     & 45\%$-$50\% &  63.1 & 50\%$-$54\% &  92.9 \cr
	 &	$-$	     & 45\%$-$55\% &  78.1 & 50\%$-$59\% &  58.5 \cr
	 &	$-$	     & 45\%$-$60\% &  83.6 & 50\%$-$63\% &  56.1 \cr
	 &	$-$	     & 45\%$-$65\% &  59.2 & 50\%$-$66\% &  29.1 \cr
	 &	$-$	     & 45\%$-$70\% &  52.6 & 50\%$-$70\% &  16.6 \cr
	 &	$-$	     & 45\%$-$75\% &  55.1 & 50\%$-$75\% &  7.7 \cr
	 &	$-$	     & 50\%$-$55\% &  87.3 & 54\%$-$59\% &  32.7 \cr
	 &	$-$	     & 50\%$-$60\% &  83.3 & 54\%$-$63\% &  37.8 \cr
	 &	$-$	     & 50\%$-$65\% &  77.9 & 54\%$-$66\% &  13.4 \cr
	 &	$-$	     & 50\%$-$70\% &  66.5 & 54\%$-$70\% &  6.5 \cr
	 &	$-$	     & 50\%$-$75\% &  68.7 & 54\%$-$75\% &  3.2 \cr
	 &	$-$	     & 55\%$-$60\% &  87.9 & 59\%$-$63\% &  78.6 \cr
	 &	$-$	     & 55\%$-$65\% &  55.1 & 59\%$-$66\% &  18.6 \cr
	 &	$-$	     & 55\%$-$70\% &  45.0 & 59\%$-$70\% &  8.0 \cr
	 &	$-$	     & 55\%$-$75\% &  49.9 & 59\%$-$75\% &  3.4 \cr
	 &	$-$	     & 60\%$-$65\% &  33.4 & 63\%$-$66\% &  7.2 \cr
	 &	$-$	     & 60\%$-$70\% &  30.5 & 63\%$-$70\% &  3.3 \cr
	 &	$-$	     & 60\%$-$75\% &  36.6 & 63\%$-$75\% &  1.5 \cr
	 &	$-$	     & 65\%$-$70\% &  59.9 & 66\%$-$70\% &  14.0 \cr
	 &	$-$	     &	65\%$-$75\% &  66.5 & 66\%$-$75\% &  4.1 \cr
	 &	$-$	     & 70\%$-$75\% &  95.7 & 70\%$-$75\% &  12.2 \cr
\noalign{\vskip 5pt\hrule\vskip 3pt}}}
\endPlancktablewide
\endgroup
\end{table*}

For the baseline mask, in Tab.~\ref{tab:params_ext}, we report the constraints on $\tau$, $\ln(10^{10} A_{\rm s})$ both with $r=0$ and variable $r$ from the low-multipole dataset alone, having fixed the other $\Lambda$CDM parameters to the best fit of \citet{Pagano:2019tci}. We leave to the next section a detailed discussion about the $\tau$ constraints and the consequences for the cosmological scenario.

 \begin{table}[htbp!]
\begingroup
\caption{Constraints on $\ln(10^{10}A_{\rm s})$, $\tau$, and $r$ from the \combdata\ likelihood. We show mean and 68\,\% confidence levels. For $r$, the 95\,\% upper limit is shown.}
\label{tab:params_ext}
\nointerlineskip
\vskip -3mm
\setbox\tablebox=\vbox{
   \newdimen\digitwidth 
   \setbox0=\hbox{\rm 0} 
   \digitwidth=\wd0 
   \catcode`*=\active 
   \def*{\kern\digitwidth}
   \newdimen\signwidth 
   \setbox0=\hbox{+} 
   \signwidth=\wd0 
   \catcode`!=\active 
   \def!{\kern\signwidth}
\halign{\hbox to 0.9in{#\leaderfil}\tabskip=1em&
  \hfil#\hfil\tabskip=10pt& 
  \hfil#\hfil\tabskip=0pt\cr
\noalign{\doubleline}
\omit\hfil Parameter \hfil& $\Lambda$CDM & $\Lambda$CDM + $r$ \cr
\noalign{\vskip 5pt\hrule\vskip 5pt}
$\ln(10^{10} A_{\rm s})$& $ 2.978\pm0.050$& $2.82^{+0.15}_{-0.08}$\cr
\noalign{\vskip 2pt}
$\tau$& $0.069^{+0.011}_{-0.012}$& $0.067^{+0.011}_{-0.012}$\cr
$r_{0.002}$& \dots& $\leq0.79$\cr
$10^9A_{\rm s}e^{-2\tau}$& $1.715^{+0.081}_{-0.092}$& $1.48^{+0.20}_{-0.14}$\cr
\noalign{\vskip 4pt\hrule\vskip 3pt}}}
\endPlancktable
\endgroup
\end{table}

\section{Reionization constraints}\label{sec:cosmology}

CMB large scale polarization data provide an almost direct measurement of the optical depth to reionization, being $C^\mathrm{EE}_{\ell} \propto \tau^2$ and $C^\mathrm{TE}_{\ell} \propto \tau$ for multipoles $\ell \lesssim 20$. In this section we use the \combdata\ dataset in polarization, together with the \commander\ 2018 solution in temperature, to derive updated constraints on $\tau$ from CMB measurements at low frequencies. 

For the cosmological parameter tests presented in this paper we adopt the reionization model given in \citet{Lewis:2008}. This is the default model in \camb\footnote{https://camb.info} and it has been used for the \planck\ baseline cosmological results (TANH). In this model the phase change in the intergalactic medium from the almost completely neutral state (up to a residual ionization fraction of $10^{-4}$, remaining after recombination) to the ionized state is described as a sharp transition. The hydrogen reionization is assumed to happen simultaneously to the first reionization of helium, whereas the second reionization of helium is fixed at a redshift of $z=3.5$ and is again described as a sharp transition. This choice is motivated by expectations from quasar spectra. Nevertheless, we expect the modeling of the helium double ionization to have a minor impact on the final results, as varying the corresponding reionization redshift between 2.5 and 4.5 changes the total optical depth by less than $1\%$ \citep{planck2014-a25}. For the tests presented in this paper we do not explore different reionization models, however it has been shown in \citet{planck2016-l06} that $\tau$ constraints from latest \planck\ data have little sensitivity to the actual details of the reionization history. Furthermore, earlier claims by \citet{Heinrich:2018btc} of a mild evidence, in \planck\ 2015 LFI data, for a more complex model of the ionization fraction, with hints of early reionization, have not been confirmed by alternate analyses of the same data set (e.g. \citet{Villanueva-Domingo:2017ahx, Hazra:2017gtx, Dai:2018nce}). In particular, \citet{2018A&A...617A..96M} have shown how the significance of those findings has been likely overestimated due to the choice of unphysical priors. 

Having fixed the reionization model, first of all, we want to study the constraints from the large scales alone. Using the pixel-based likelihood framework of Sec. \ref{sec:likelihood} (\lowTEB) we only fit for $\tau$, $\ln(10^{10} A_{\rm s})$ and $r$, while keeping all the other $\Lambda$CDM parameters fixed to the bestfit values given in \cite{Pagano:2019tci}. Results are shown in Tab~\ref{tab:params_ext}, where the parameter $r$ is estimated at a scale $k=0.002\;\mathrm{Mpc^{-1}}$. The derived constraint on $\tau$ is

\be
\tau=0.069_{-0.012}^{+0.011} \qquad (68\%,\textrm{\lowTEB}),\label{eq:tau_lowTEB}
\ee
which corresponds to a $5.8\,\sigma$ detection from the low frequency CMB polarization data.

We then extend the analysis to also include data from the small scales, specifically adding the \planck\ 2018 likelihood for TT, TE, EE angular power spectra \citep{planck2016-l05}. This time we let all the six base $\Lambda$CDM parameters vary, and we sample from the space of possible cosmological parameters with a MCMC exploration using CosmoMC \citep{2002PhRvD..66j3511L}. The reionization optical depth estimated in this case is\footnote{In the following, the presence of the \lowTEB\ dataset should be always understood.}: 
\be
\tau=0.074_{-0.011}^{+0.010} \qquad  \onesig{\textrm{   TT,TE,EE}}.\label{eq:tau_TTTEEE_plus_lowTEB}
\ee
 The parameter constraints we derive for pure $\Lambda$CDM are given in Tab.~\ref{tab:params}, where for comparison purposes we also report the \planck\ 2018 baseline results. The two compared datasets differ by the low-$\ell$ likelihoods. In one case there is the pixel-based likelihood developed in this paper (\lowTEB), while in the other case the low-$\ell$ likelihood is a combination of the Blackwell-Rao estimator for the \commander\ temperature solution and the E-mode power spectrum based \Planck\ Legacy HFI likelihood (\lowE). The latter likelihood provides a constraint on $\tau$ that is about $1.5$ times tighter and $1.4\,\sigma$ lower in value than the one we obtain from the \combdata\ likelihood. Due to the well known degeneracy between $A_s$ and $\tau$, this also translates in a $33\%$ tighter constraint on $\ln(10^{10} A_{\rm s})$ with a $1.8\,\sigma$ lower value. All the other cosmological parameters, instead, are in good agreement, differing by at most $36\%$ of the $\sigma$. A similar behaviour is also found when Tab.~\ref{tab:params} is compared with an analogous analysis shown in \citet{Pagano:2019tci}.

\begin{table*}
\begingroup
\caption{Parameter constraints for the \LCDM\ cosmology \citep[as defined in][]{planck2013-p11}, illustrating the impact of replacing the low-$\ell$ baseline \Planck\ 2018 likelihood (\lowE) with the \combdata\ likelihood presented in this paper (\lowTEB). We also show the change when including the high-$\ell$ polarization likelihood in the analysis.}
\label{tab:params}
\openup 5pt
\newdimen\tblskip \tblskip=5pt
\nointerlineskip
\vskip -3mm
\normalsize
\setbox\tablebox=\vbox{
    \newdimen\digitwidth
    \setbox0=\hbox{\rm 0}
    \digitwidth=\wd0
    \catcode`"=\active
    \def"{\kern\digitwidth}
    \newdimen\signwidth
    \setbox0=\hbox{+}
    \signwidth=\wd0
    \catcode`!=\active
    \def!{\kern\signwidth}
\halign{
     \hbox to 0.9in{$#$\leaderfil}\tabskip=1.5em&
     \hfil$#$\hfil&
     \hfil$#$\hfil&\hfil$#$\hfil&
     \hfil$#$\hfil\tabskip=0pt\cr
\noalign{\doubleline}
\multispan1\hfil \hfil&\multispan1\hfil TT+\lowE\hfil&\multispan1\hfil TT+\lowTEB\hfil&\multispan1\hfil TTTEEE+\lowE\hfil&\multispan1\hfil TTTEEE+\lowTEB\hfil\cr
\noalign{\vskip -3pt}
\omit\hfil Parameter\hfil&\omit\hfil 68\,\% limits\hfil&\omit\hfil 68\,\% limits\hfil&\omit\hfil 68\,\% limits\hfil&\omit\hfil 68\,\% limits\hfil\cr
\noalign{\vskip 3pt\hrule\vskip 5pt}
\Omega_{\mathrm{b}} h^2&0.02212\pm 0.00022&0.02218\pm 0.00022&0.02236\pm 0.00015&0.02241\pm 0.00015\cr
\Omega_{\mathrm{c}} h^2&0.1206\pm 0.0021&0.1200\pm 0.0021&0.1202\pm 0.0014&0.1197\pm 0.0014\cr
100\theta_{\mathrm{MC}}&1.04077\pm 0.00047&1.04086\pm 0.00047&1.04090\pm 0.00031&1.04096\pm 0.00031\cr
\tau&0.0522\pm 0.0080&0.071_{-0.011}^{+0.010}&0.0544^{+0.0070}_{-0.0081}&0.074_{-0.011}^{+0.010}\cr
\ln(10^{10} A_\mathrm{s})&3.040\pm 0.016&3.076\pm 0.021&3.045\pm 0.016&3.082\pm 0.021\cr
n_\mathrm{s}&0.9626\pm 0.0057&0.9645\pm 0.0058&0.9649\pm 0.0044&0.9664\pm 0.0044\cr
\noalign{\vskip 3pt\hrule\vskip 5pt}
H_0&66.88\pm 0.92&67.12\pm 0.93&67.27\pm 0.60&67.51\pm 0.61\cr
\Omega_{\mathrm{m}}&0.321\pm 0.013&0.317\pm 0.014&0.3166\pm 0.0084&0.3134\pm 0.0084\cr
\Omega_{\mathrm{\Lambda}}&0.679\pm 0.013&0.683\pm 0.013&0.6834\pm 0.0084&0.6866\pm 0.0084\cr
\sigma_8&0.8118\pm 0.0089&0.825\pm 0.010&0.8120\pm 0.0073&0.8259\pm 0.0091\cr
z_{\mathrm{re}}&7.50\pm0.82&9.3\pm 1.0&7.68\pm0.79&9.51_{-0.97}^{+0.98}\cr
10^9 A_{\mathrm{s}} &2.092\pm 0.034&2.167_{-0.049}^{+0.043}&2.101^{+0.031}_{-0.034}&2.181_{-0.049}^{+0.043}\cr
10^9 A_{\mathrm{s}} e^{-2\tau}&1.884\pm 0.014&1.882\pm 0.014&1.884\pm 0.012&1.882\pm 0.012\cr
\mathrm{Age}/\mathrm{Gyr}&13.830\pm 0.037&13.819\pm 0.037&13.800\pm 0.024&13.791\pm 0.024\cr
\noalign{\vskip 5pt\hrule\vskip 3pt}
} 
} 
\endPlancktable
\endgroup
\end{table*}

Comparing constraints in Tab.~\ref{tab:params_ext} and Tab.~\ref{tab:params}, we note that the values of $\ln(10^{10} A_{\rm s})$ and $\tau$ derived from the large scales alone are $1.9\,\sigma$ and $0.4\,\sigma$ lower, respectively. This behaviour has been firstly noticed in \citet{planck2014-a13}, and it is induced by the low-$\ell$ \textit{anomaly}, i.e. the power deficit in the measured $TT$ power spectrum with respect to the best fit model at multipoles between $\ell =20$ and $30$. When limiting the analysis to the large scales, that is to multipoles up to $30$, the deficit has a high relative weight in the final solution leading to a value of the overall amplitude of the spectrum that is lower than the one from the full analysis that includes multipoles up to $\ell =2500$. Due to the aforementioned degeneracy, this also results in a lower value of $\tau$. 

\begin{figure}[h]
	\includegraphics[width=0.45\textwidth]{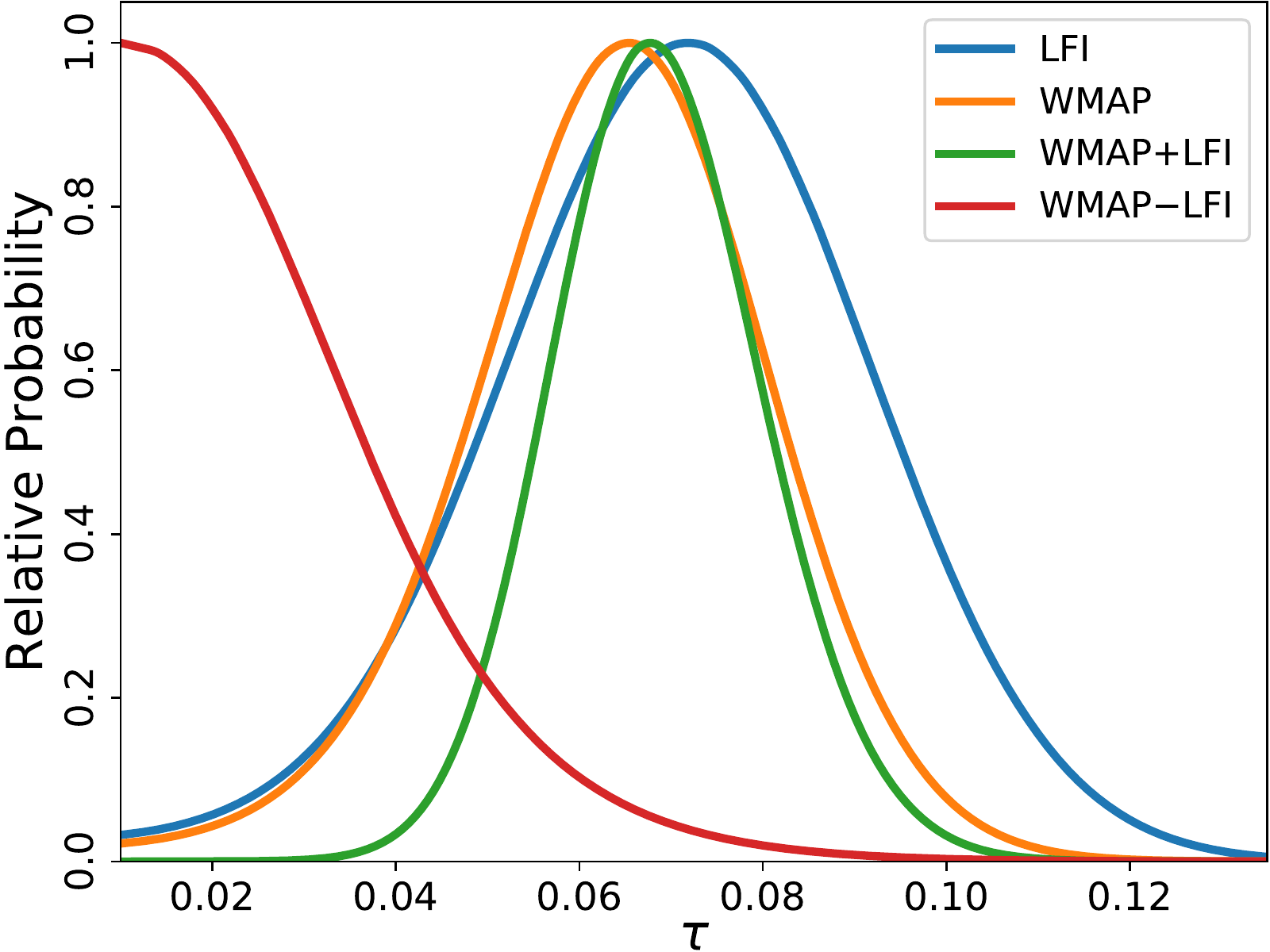}
	\caption{Posterior distributions of $\tau$. LFI (blue) and \wmap\ (orange) are computed on their own 50\% mask, \combdata\ (green) in the union of the two, retaining 54\% of the sky, while \wmap$-$LFI (red) is computed in their intersection, retaining 46\% of the sky.\label{fig:tau_ptau_all}}
\end{figure}

Since one of the main results of this paper is the $\tau$ constraint from the \combdata\ dataset, we want to further comment on the robustness of this result. In Fig.~\ref{fig:tau_ptau_all} we show the good agreement between the estimates of $\tau$ from LFI and \wmap\ separately. The two have been derived using their own $f_{sky}=50\%$ mask. The consistency between the two experiments is further confirmed by the null test that we perform estimating $\tau$ from the half-difference map of the two data sets, LFI$-$\wmap. The posterior distribution for this case is reported in the same figure and it is compatible with noise, giving an upper limit of $\tau \leq 0.059$ at $95\%$ CL.

\begin{figure}[h]
	\includegraphics[width=0.45\textwidth]{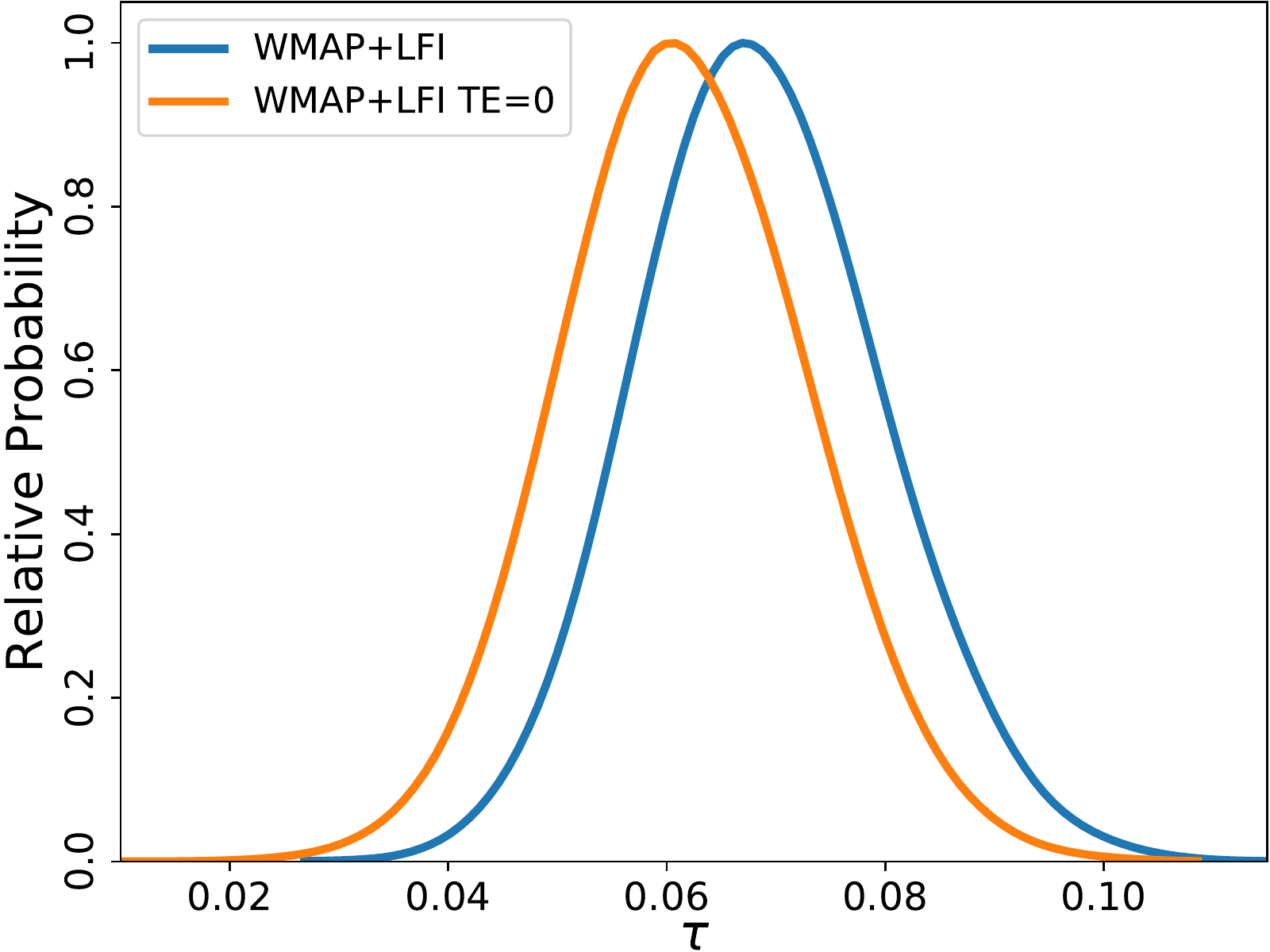}
	\caption{Comparison of $\tau$ posterior distributions for the \combdata\ using the full TQU likelihood (blue) and imposing $TE=0$ in order to factorize the T and QU parts of the likelihood (orange).\label{fig:tau_ptau_noTE}}
\end{figure}

Differently from the baseline low-$\ell$ \planck\ 2018 likelihood, which is based on the $TT$, $EE$ and $BB$ power spectra, the pixel-based likelihood used in this paper also includes the information contained in the $TE$ cross power spectrum. In order to investigate the impact of this extra information, we build a polarization-only version of the pixel-based likelihood, which contains only the Q and U maps and the QQ, QU, and UU blocks of the covariance matrix, in analogy of what done in \citet{planck2014-a13}. When we use this latter likelihood we employ low-$\ell$ $TT$ \commander\ likelihood based on Blackwell-Rao estimator. The value of $\tau$ measured nulling the $TE$ cross correlation is

\be
\tau=0.062\pm0.012 \qquad (68\%,\textrm{\commander\ + \lowEB}),\label{eq:tau_lowTEBnoTE}
\ee

which represents roughly a half-$\sigma$ downward shift with respect to the full TEB likelihood, already seen on the LFI only likelihood in \citet{planck2014-a13}. Such behavior is also shown by \wmap\ which, on 50\% sky, yields $\tau=0.055_{-0.017}^{+0.019}$ forcing $TE=0$ and $\tau=0.064_{-0.015}^{+0.017}$ with the full TEB likelihood. For \wmap\ the same behavior is also present on larger masks, for example on the 75\% sky fraction we measure $\tau=0.065_{-0.014}^{+0.013}$ when $TE=0$ and $\tau=0.070_{-0.013}^{+0.012}$ when also TE is varied.\\
In all the previous cases, when TE is forced to zero, the $\tau$ posterior shifts closer to the HFI determination \citep{planck2016-l05,Pagano:2019tci} which is based only on $EE$ estimates. Posteriors of the full pixel-based likelihood and the one without $TE$ for \combdata\ are shown in Fig.~\ref{fig:tau_ptau_noTE}. 

\begin{figure}[h]
	\includegraphics[width=0.45\textwidth]{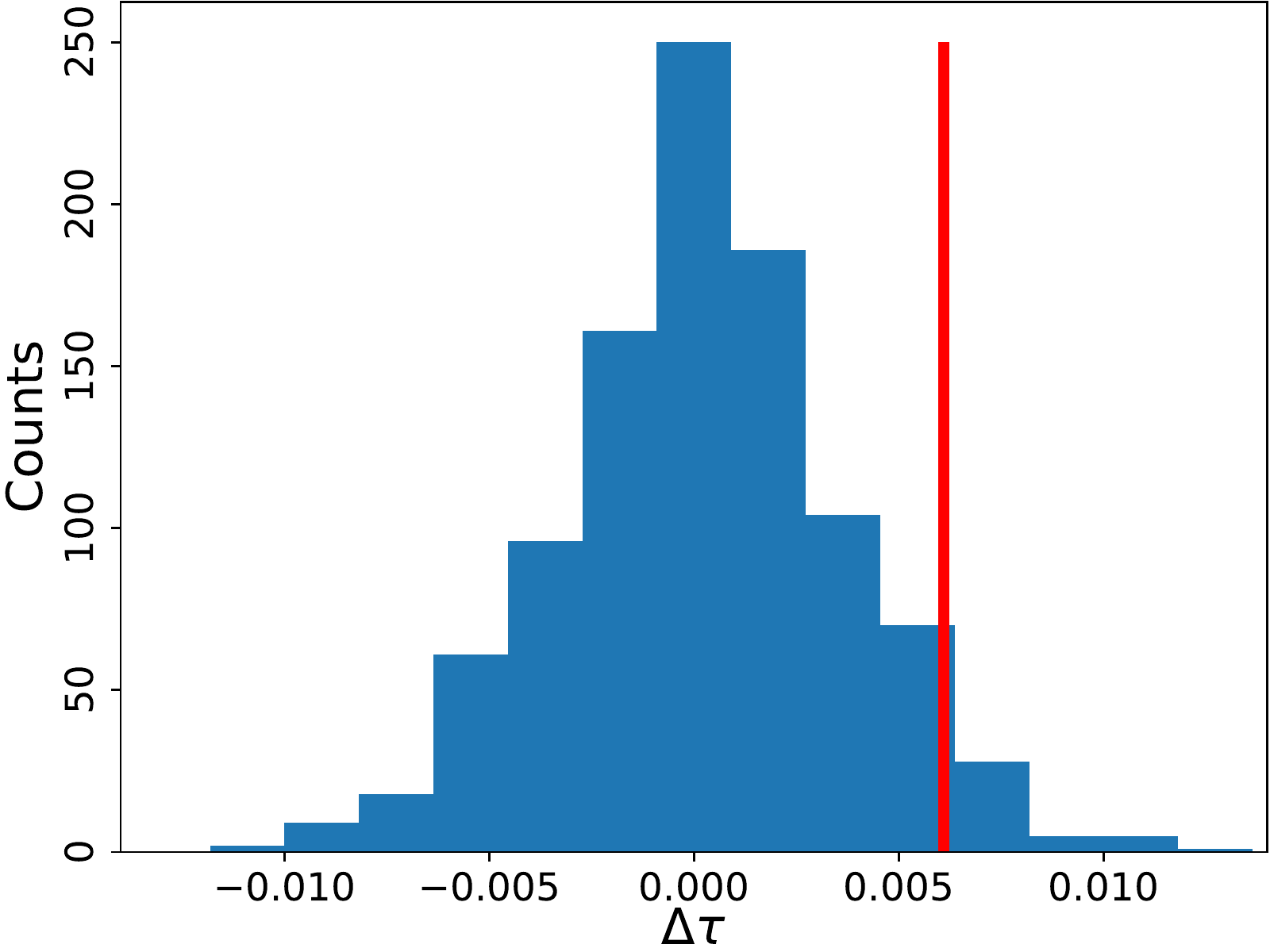}
	\caption{Histogram of $\Delta\tau\equiv\tau_{\rm Full}-\tau_{\rm noTE}$ obtained analyzing a Montecarlo of 1000 simulations. The red vertical bar shows value measured on data.\label{fig:tau_with_and_withoutTE_validation_delta}}
\end{figure}

\begin{figure}[h]
	\includegraphics[width=0.45\textwidth]{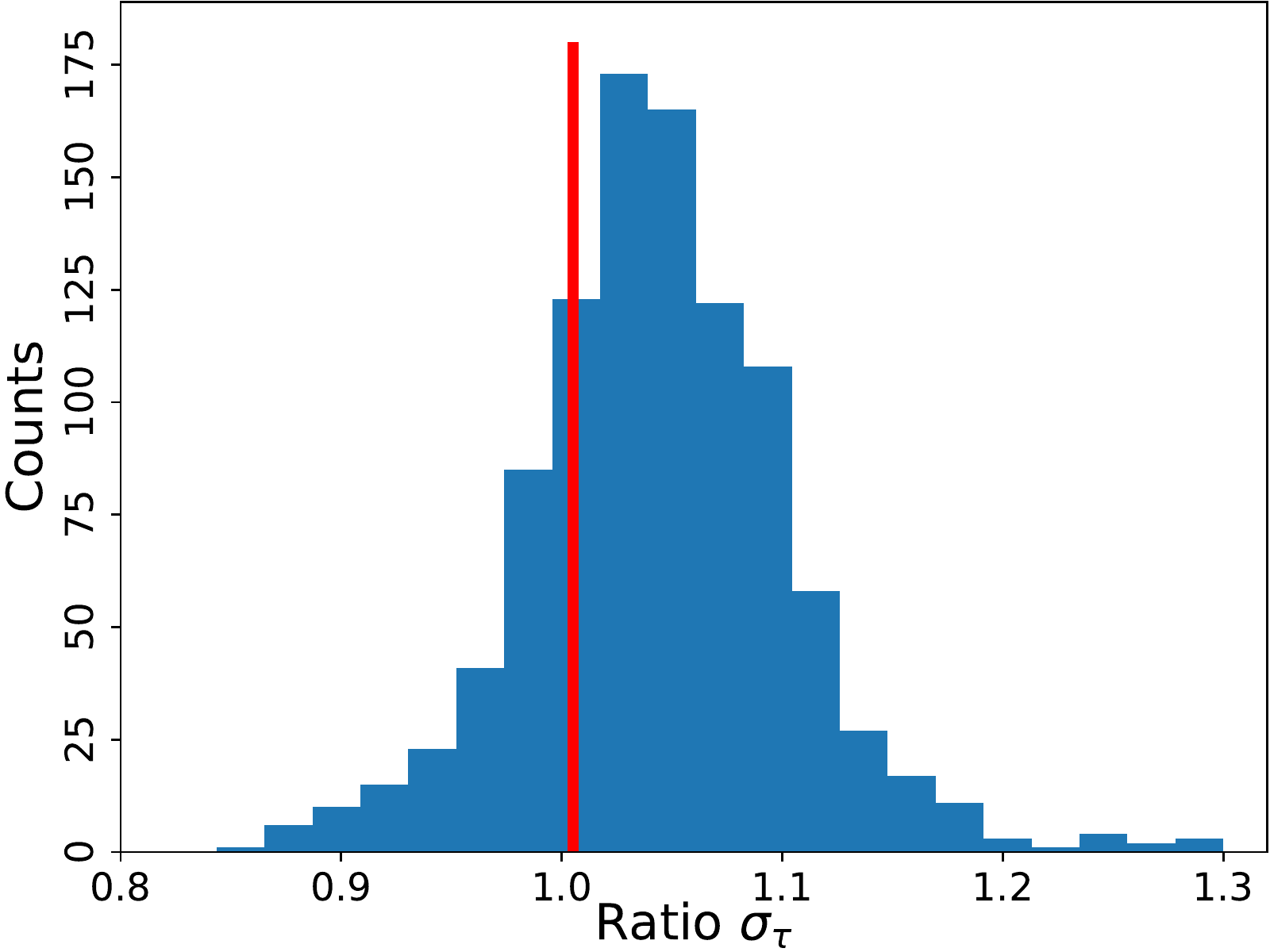}
	\caption{Histogram of $\sigma_{\tau_{\rm noTE}}/\sigma_{\tau_{\rm Full}}$ obtained analyzing a Montecarlo of 1000 simulations. The red vertical bar shows value measured on data.\label{fig:tau_with_and_withoutTE_validation_sigma}}
\end{figure}

In order to verify if such behaviour is coherent with our error budget, we compare the shift in $\tau$ with a set of simulations. In Figure~\ref{fig:tau_with_and_withoutTE_validation_delta} we show the histogram of $\Delta\tau$, defined as the difference between the $\tau$ estimated form the full TEB likelihood (``Full'') and the $\tau$ estimated forcing $TE=0$ (``noTE''), for a Montecarlo of 1000 signal and noise simulations. This analysis shows that nullifying $TE$ still provides an unbiased estimation of $\tau$ and also that the shift observed in data is not anomalous, representing a $1.7~\sigma$ fluctuation. We also show in Fig.~\ref{fig:tau_with_and_withoutTE_validation_sigma}  a similar plot for the ratio of 1-$\sigma$ errors defined as $\sigma_{\tau_{\rm noTE}}/\sigma_{\tau_{\rm Full}}$, also in this case the value measured on data is compatible with simulations. This test also suggests that, for this dataset, removing TE degrades $\sigma_{\tau}$ by about $5\%$ on average.

Adding to the CMB temperature and polarization data the \Planck\ lensing likelihood \citep{planck2016-l08} and baryon acoustic oscillation (BAO) measurements \citep{Alam:2016hwk,Beutler:2011hx,Ross:2014qpa} breaks more efficiently the degeneracy with the amplitude of the scalar perturbations providing 

\be
\tau=0.0714_{-0.0096}^{+0.0087} \qquad \onesig{\textrm{   TT,TE,EE+Lensing+BAO}}. \label{eq:tau_TTTEEE_plus_lowTEB_plus_lensing_plus_BAO}
\ee

Assuming the TANH model for the ionzation fraction the $\tau$ constrain can be directly converted into a mid-point reionization redshift of

\be
z_{\rm re}=9.28\pm0.84 \qquad \onesig{\textrm{   TT,TE,EE+Lensing+BAO}}. \label{eq:tau_TTTEEE_plus_lowTEB_plus_lensing_plus_BAO}
\ee

This value is higher but still compatible with analogous estimates that use instead the \Planck\ HFI based large scale polarization likelihood, $z_{\rm re}=7.82 \pm 0.71$ \citep{planck2016-l06} and $z_{\rm re}=8.21 \pm 0.58$ \citep{Pagano:2019tci}.

The \combdata\ CMB map and the corresponding covariance matrix are packaged in low-$\ell$ likelihood modules compatible with the \clik\ infrastructure \citep{planck2013-p08,planck2013-p28,planck2014-ES,planck2016-ES} which are made publicly available\footnote{The \combdata\ likelihood module is available on \href{https://web.fe.infn.it/~pagano/low\_ell\_datasets/wmap\_lfi\_legacy}{https://web.fe.infn.it/$\sim$pagano/low\_ell\_datasets/wmap\_lfi\_legacy}}. We provide both a likelihood module that implements Eq.~\ref{eq:multivariate_gaussian} inverting the full covariance matrix and one that implements the Sherman-Morrison-Woodbury (SMW) formula \citep{GoluVanl96} which allows to speed up by an order of magnitude the computation \citep[see][Appendix B.1 for details]{planck2014-a13}, but does not include TB an EB. In both cases, in order to keep full compatibility with the codes of clik framework, we do not treat the regularization noise as described in sec.~\ref{sec:methods}, but we sum instead a single realization. Such noise realization has been chosen in order to have a deviation for $\ln(10^{10} A_\mathrm{s})$ and $\tau$ with respect to the baseline case smaller than $1\%$ in units of $\sigma$.

\section{Conclusions \label{sec:concl}}

We have presented a novel CMB pixel-space likelihood focussed on polarization at large angular scales, whose main cosmological target is the optical depth to reionization $\tau$. The underlying dataset combines foreground-mitigated \wmap\ Ka, Q and V bands with \Planck\ LFI 70 GHz channel in an optimally weighted CMB map. In the foreground cleaning of \wmap\ bands we adopt the \planck\ 353 \GHz\ channel, as dust template, instead of the \wmap\ dust model based on starlight-derived polarization directions. As a synchrotron template, K band is used for \wmap\ channels, while \planck\ 30 GHz for the 70 GHz map. The corresponding covariance matrix is computed coherently and fed, together with the cleaned CMB map, into a pixel space likelihood, made publicly available with this paper. A set of masks with increasing sky fraction has been produced and used to test the performance of the component separation, the quality of polarization power spectra, and the overall stability of $\tau$ constraints, showing a remarkable stability among sky fractions.

For the baseline dataset, which retains 54\% of the sky, the $\ell$-by-$\ell$ probability to exceed the $\chi^2$ of the measured angular power spectra (PTE) does not show any major outlier, with only $\ell = 18$ and $23$ of BB and $\ell=23$ of EB spectra at more than 2.5-$\sigma$. Consequently, the integrated PTEs are perfectly consistent with simulations both on the reionization peak only (i.e. $\ell=2\div10$) and on the full multipole range (i.e. $\ell=2\div29$).

Regarding the reionization optical depth estimation, we compared the variation of $\tau$ estimated on different sky fractions with a Montecarlo of signal plus noise, finding no significant deviations for the baseline dataset compared with other sky fractions, up to $f_{\rm sky}\sim70\%$.

\begin{figure}[h]
	\includegraphics[width=0.45\textwidth]{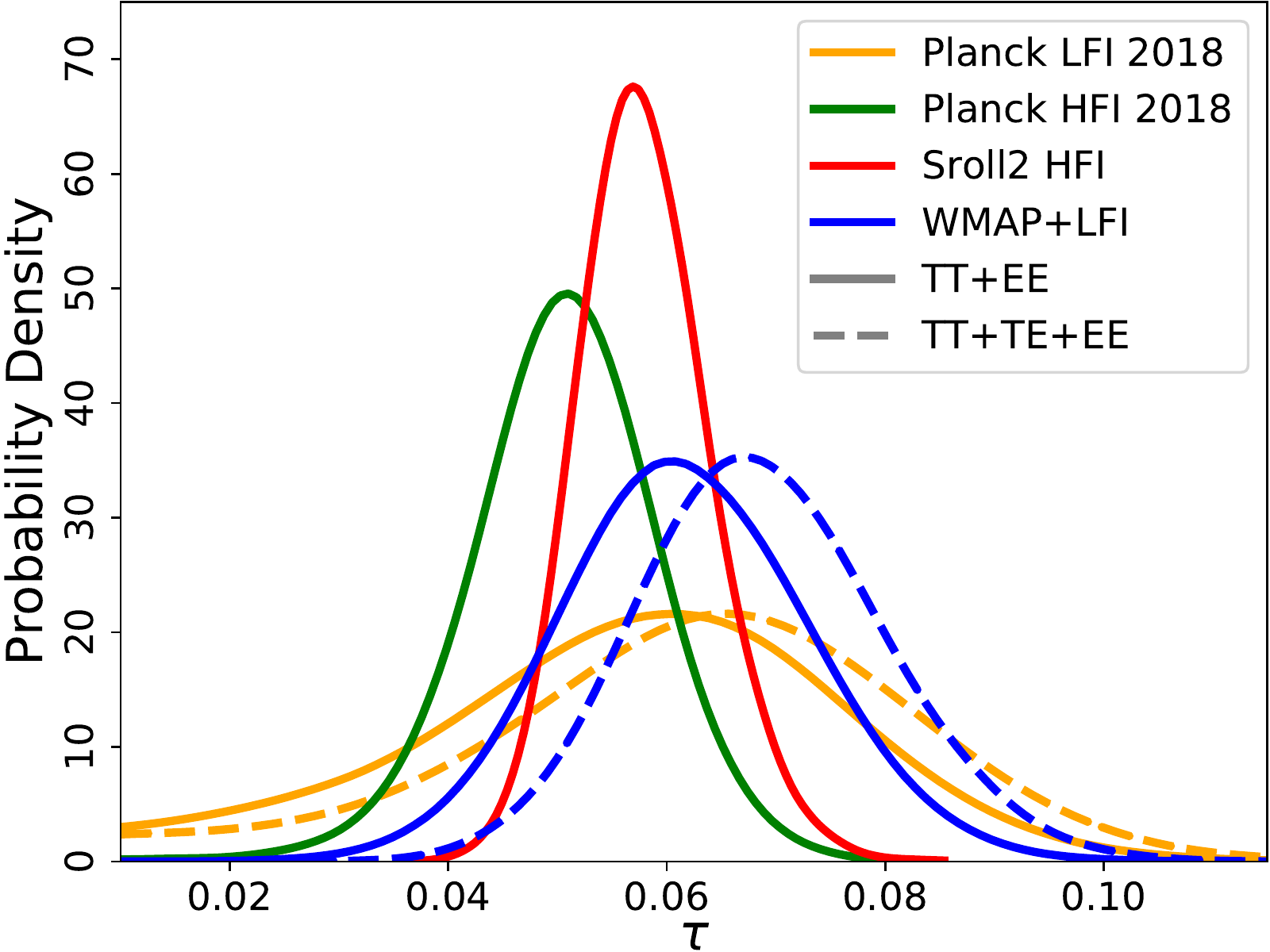}
	\caption{Posterior distributions for $\tau$ for various datasets. Solid lines represent $\tau$ constraints ignoring $TE$, dashed lines assume full TEB likelihood. Only $\tau$ and $\ln(10^{10} A_\mathrm{s})$ are sampled, the remaining $\Lambda$CDM parameters are fixed to common fiducial values. The green and yellow lines show results obtained with the official $\planck$ Legacy low-$\ell$ likelihoods \citep{planck2016-l05}. The red line is obtained running the \srolltwo\ likelihood \citep{Pagano:2019tci}. The blue lines represent the result of this paper.\label{fig:tau_all_datasets}}
\end{figure}

Sampling the parameter space with our low-$\ell$ likelihood only, we find $\tau=0.069^{+0.012}_{-0.011}$. When CMB small scales, BAO observations and \Planck\ lensing likelihood are included we shrink optical depth constraint down to $\tau=0.0714_{-0.0096}^{+0.0087}$. Such bounds are slightly less constraining when compared with the existing \Planck\ HFI based likelihood \citep[see e.g.][and Fig. \ref{fig:tau_all_datasets}]{planck2016-l05,Pagano:2019tci}, yet represent a novel measure obtained with an independent pipeline which adopts different data and likelihood approximation and includes TE correlation, while the \Planck\ HFI estimates are currently restricted to EE information. The $\tau$ estimates obtained with the likelihood package discussed in this paper is in general well compatible with the \Planck\ HFI based constraints, with a preference for slightly higher $\tau$ values, probably driven by the inclusion of TE. This bound is also in perfect agreement with the \Planck\ LFI Legacy likelihood \citep{planck2016-l05}. Within the $\Lambda$CDM model $\tau$ can be also constrained  without the use of polarization data  breaking the degeneracy between $A_s$ and $\tau$ combining temperature and weak lensing data. In this kind of analysis the \Planck\ collaboration found $\tau=0.080\pm0.025$ from temperature and lensing data, and $\tau=0.078\pm0.016$ when also BAO is added \citep{planck2016-ES}. Those values, despite being slightly higher than previous findings \citep{Weiland:2018kon,planck2014-a15}, are still in agreement with our constraints. 

The likelihood package we provide is built on a real space estimator and does not assume rotational invariance, but only Gaussianity of the fields, which secures a few advantages. For instance, it can also be easily used for constraining non-rotationally invariant cosmologies, including naturally the TB and EB spectra in the parameter exploration.  

\begin{acknowledgements}
We acknowledge the use of \camb, \healpix\ and Healpy software packages, and the use of computing facilities at CINECA. We are grateful to M. Gerbino for useful suggestions. MM is supported by the program for young researchers ``Rita Levi Montalcini" year 2015. We acknowledge support from the COSMOS network (www.cosmosnet.it) through the ASI (Italian Space Agency) Grants 2016-24-H.0 and 2016-24-H.1-2018. 
 \end{acknowledgements}

\bibliographystyle{aat}

\interlinepenalty=10000

\bibliography{low_ell_lfi_likelihood,Planck_bib}

\end{document}